%% file: wsc21.tex
\documentclass{wscpaperproc}
\usepackage{latexsym}
\usepackage{graphicx}
\usepackage{mathptmx}
\usepackage{subcaption}
\usepackage{amsmath}
\usepackage{amsfonts}
\usepackage{amssymb}
\usepackage{amsbsy}
\usepackage{amsthm}
\usepackage{soul}
\usepackage{color}

\usepackage{hyperref}
\hypersetup{colorlinks=true,urlcolor=blue,citecolor=black,anchorcolor=black,linkcolor=black}

\newtheoremstyle{wsc}
{3pt}
{3pt}
{}
{}
{\bf}
{}
{.5em}
{}

\theoremstyle{wsc}

\newtheorem{definition}{Definition}

\begin{document}

%
%

\pagestyle{fancyplain}

\thispagestyle{plain}
\firstPageHead{}

\chead{\fancyplain{}{\itshape Streitmatter and Zhang}}

\rhead{}
\cfoot{}
\renewcommand{\headrulewidth}{0pt} 

\input{wscbib.tex}           

\setlength{\baselineskip}{12.7pt}

\title{Travel Cadence and Epidemic Spread}

\author{Lauren Streitmatter\\
Division of Engineering Science\\
Faculty of Applied Science and Engineering\\
University of Toronto\\
40 St George St\\
Toronto, ON M5S 2E4, CANADA
\and
Peter Zhang\\
Heinz College of Information Systems and Public Policy\\
Carnegie Mellon University\\
4800 Forbes Ave\\
Pittsburgh, PA 15213, USA
}

\maketitle

\vspace{-0.2in}
\section*{ABSTRACT}
In this paper, we study how interactions between populations impact epidemic spread. We extend the classical SEIR model to include both integration-based disease transmission simulation and population flow. Our model differs from existing ones by having a more detailed representation of travel patterns, without losing tractability. This allows us to study the epidemic consequence of inter-regional travel with high fidelity. In particular, we define \emph{travel cadence} as a two-dimensional measure of inter-regional travel, and show that both dimensions modulate epidemic spread. This technical insight leads to policy recommendations, pointing to a family of simple policy trajectories that can effectively curb epidemic spread while maintaining a basic level of mobility.

\section{INTRODUCTION}
\label{sec:intro}

In this paper, we study how interactions between populations impact epidemic spread. This is an important and classical question that the simulation community can  provide insights with for two reasons. First, such impact should ideally be understood before a disease outbreak. Modern disease outbreaks tend to originate from novel sources, and it is impractical to study such effects to the fullest extent with minimal clinical or public health data during early stages of the spread. Therefore, simulation should play an important role. Second, the complexity and scale of travel dynamics have been continuously evolving in the past few decades on the global, national, and local levels. Thus, classical disease transmission methods in simulation should be updated for the new context, to address large scale, high fidelity, and dynamic problems.

To answer the question of how interactions between populations impact epidemic spread, we first design a computationally efficient tool (Section \ref{sec:method}) to capture population dynamics both within a community and across communities, incorporating a flexible and detailed modeling on travel patterns. This allows us to take advantage of the computational tractability of integration-based simulation models and the fidelity of agent-based models. With this tool, we are then able to characterize how travel patterns impact disease transmission (Section \ref{sec:analysis}).

\subsection{Literature Review}
\label{subsec:lit}

Before introducing our method and analysis in detail, it is important to visit the literature on disease transmission involving travel dynamics. We do not intend to provide an exhaustive review, because there is a vast literature, and it is impossible to do a fair job if one's goal is to be thorough. Rather, we rely on our own understanding of the conceptual difference between different lines of research, point out why certain gaps exist, and discuss how we take steps to address such gaps.

In the classical Susceptible-Exposed-Infectious-Recovered (SEIR) model, individuals are assumed to be in one of four states. When a susceptible person interacts with an exposed or infected person, there is a chance that the susceptible person transitions into the exposed state.
A key assumption that enables model tractability is that interactions are uniform -- the probability distribution over pairwise encounters is uniform. It is possible to relax this assumption to some extent, but one's behavior is still dictated by its state, not by location or travel specific characteristics.

Several streams of research have already looked at adding population mixing or traveling beyond the simple SEIR setting. Our model differs from existing approaches by having a more detailed modeling of travel for a longer term. In the remaining part of this subsection, we briefly review these studies.

Two streams of research are closest to our work. 
The first stream studies the impact of travel behavior on the (short term) epidemic spread, usually with real world data and employing empirical methods. The effect of travel on disease transmission has become an intensely studied subject since the spread of COVID-19. Given the heterogeneity in travel behavior across different countries, and given the availability of data worldwide, researchers naturally tried to understand the effect of travel on disease transmission by applying empirical methods. 
For example, \shortciteN{Chinazzi2020}, \shortciteN{Costantino2020}, \shortciteN{Adekunle2020}, and \shortciteN{Anzai2020} study the effect of travel ban/restriction on the delaying of disease spread across countries. Some other projects also study such effects on a regional scale \shortcite{Mangrum2020}. However, these studies focus on the effect of travel pattern change at one time point. They do not study the effect of long term travel behavior change, or the long term effect of a single travel behavior change. Our model and analysis are aimed at addressing this gap.

The second stream focuses on simulation-based and analytical understanding of the long term impact of travel on disease transmission, under the name of \emph{metapopulation disease transmission models} or variants thereof. \shortciteN{Ball2015} provides an excellent review and critical analysis of this stream of works. Overall, these metapopulation models focus on why and how subpopulations (compartments) with heterogeneous characteristics have different disease transmission trajectories over time \shortcite{Watts2005,Ni2009,Apolloni2014}. 
Traveling between compartments is also modelled, but usually done in the ambience: there is usually a set of parameters to model travel likelihood between any two compartments. Such parameters indirectly control the actual population flow, and usually remain the same over time. In this current project, our focus is to divide traveling into multiple episodes, each one with a different set of travel parameters; and each set of the parameters would model the amount of population flow precisely and directly. 

A few other streams of research are also related, including network-type disease transmission models and control theory perspectives. Network-type models delegate the description of travel behavior to edge-weights in a network, which usually has limited fidelity in modeling the complex interaction behavior between different nodes. Control theory type of disease transmission models usually prefer to keep the model setup pristine and not model travel details, in order to generate clean analytical results. Under the control theory lens, both descriptive \shortcite{Arino2003} and prescriptive \shortcite{Birge2020} models have been proposed.

\section{METHODOLOGY}
\label{sec:method}

In this section, we describe our epidemic simulation model design in detail. We note that this model is not just for the analysis of our travel cadence question, but a general purpose model that simulates SEIR trajectories with partitions and population flows. Therefore, this section pertains to the general design of our simulation model. For the model customization that supports the analysis on travel cadence and epidemic spread, we wait until Section \ref{sec:analysis}.

We coded the model in Python in a Jupyter Notebook. All of the details included in the following sections are already incorporated into the code and are fully functional. The code is shown here. 

\subsection{Dynamic SEIR Model}
The SEIR (Susceptible, Exposed, Infected, Recovered) differential equations below are used as the foundation for this equation-based model. Terms are included to account for the percentage of infected individuals who quarantine themselves ($q$) and to account for the delay between exposure and symptom onset ($\alpha$).

\vspace{-0.2in}
\begin{align}\label{eq:SEIR}
  & S'(t) = -\beta S (1-q)I/N- \alpha S E/N\\
  & E'(t) = \beta S (1-q)I/N + \alpha S E/N - \sigma E\\
  & I'(t) = \sigma E - \gamma I \\
  & R'(t) = \gamma I
\end{align}

\begin{figure}[h!]
    \centering
    \includegraphics[width=0.6\linewidth]{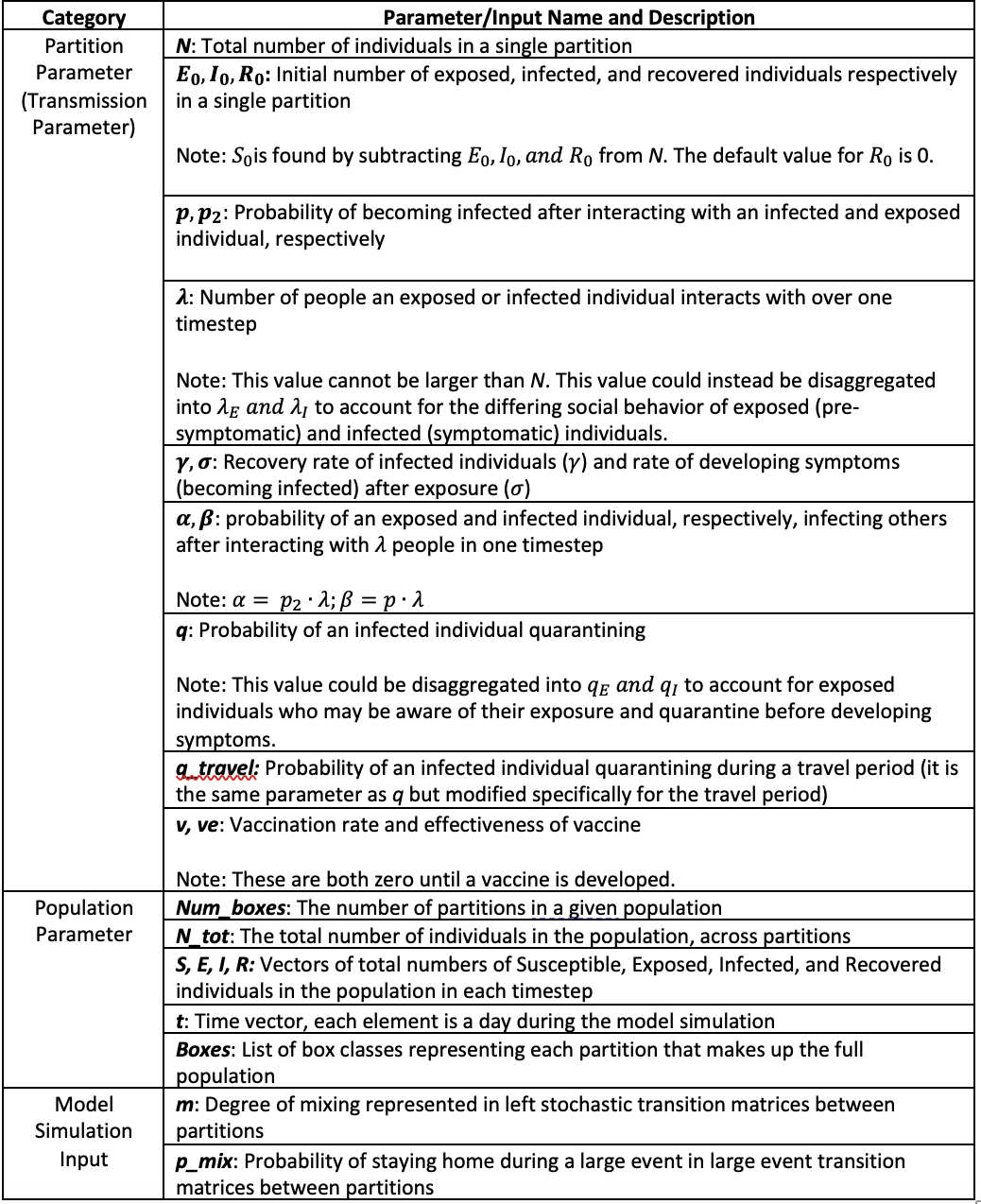}
    \caption{Definitions and notations.}
    \label{fig:Parameter_Table}
\end{figure}

The list of parameters in Figure \ref{fig:Parameter_Table} describes each variable included in the model. At any time throughout a simulation, it is possible to update one or more of the partitions' parameter values to reflect the dynamic nature of disease spread and transmissibility.

\subsection{Population Partitioning}
This model extends beyond the classic SEIR model described above by allowing a total population of size $N_{\text{tot}}$ to be split into \texttt{Num\_boxes} compartments or partitions (we use the word partition here to refer to a single compartment, different from the set-theoretic definition of partition, which refers to the collection of all compartments). Each partition is isolated with its own $S$, $E$, $I$, and $R$ groups that evolve independently from the other partitions depending on its local parameters. This can be used to represent many real-life situations and strategies, including geographical separation across different neighborhoods, counties, or states or designing school class schedules to reduce interaction between ``pods''.

To study a classical SEIR model rather than a partitioned model, the number of partitions can be set to one. However, the partition feature of this model allows populations to be disaggregated into smaller groups without the drastic increase in computational power from an agent-based rather than equation-based model.
To set up the model, there are two different methods:
\begin{enumerate}
    \item Initialize population with uniform parameters across partitions.
    \item Initialize population with a list of initialized partitions that comprise it.
\end{enumerate}

Method 1 initializes a population of size $N_{\text{tot}}$ with \texttt{Num\_boxes} partitions, each partition assigned uniform sub-population sizes and transmission parameters shown in Figure \ref{fig:Parameter_Table}, excluding $S_0$, $E_0$, $I_0$, and $R_0$. This method assumes the specific locations of each individual who is initially exposed or infected is unknown, so it randomly distributes the total $E_0$  and $I_0$ in whole numbers across the different partitions so that each partition $n$ has its own $S_{0,n}$, $E_{0,n}$, $I_{0,n}$, and $R_{0,n}$ and $\sum_n S_{0,n} = S_0, \sum_n E_{0,n} = E_0, \sum_n I_{0,n} =I_0, \sum_n R_{0,n} =R_0$. 

Method 2 initializes the population from a pre-existing list of instantiated partitions. This allows each partition to have unique transmission parameters and population sizes if desired. This also allows for non-randomized control over the distribution of $E_0$  and $I_0$ across partitions if it is desirable to model scenarios where initial conditions are known in detail.

The disease spreads through each partition according to equations 1-4 above, where in partition $n$, $S_n,E_n,I_n$,and $R_n$ are used along with the partition-specific parameter values. While it is helpful to see how the disease evolves when partitions are completely separated from one another throughout the entire period, this is an ideal case that is often not achievable in reality. Instead, there is typically mixing between pods, neighborhoods, or states that this model can account for as described in the following section.

\subsection{Population Mixing}
The time evolution of $S_n, E_n, I_n$, and $R_n$ in partition $n$ are stored in $S_n,E_n,I_n$,and $R_n$ vectors of length $T$, where $T$ is the number of days in the simulation. $S_n,E_n,I_n$,and $R_n$ values for each partition at the current time $t$ are stored in size \texttt{Num\_boxes} $S,E,I$, and $R$ vectors. Thus, population mixing (when people from different partitions come into contact with each other) is represented with matrix-vector multiplication using the latter set of vectors. This is depicted graphically in Figure \ref{fig:mixing} below. We provide a high level summary of the mixing procedure here, followed by a detailed description for each step.

\vspace{-0.2in}
\begin{enumerate}
    \item Generate an \texttt{Num\_boxes} $\times$ \texttt{Num\_boxes} mixing matrix $M$ for each group ($S, E, I, $ and $R$).
	\item Calculate the unmixing matrix $M'$ using $M$ and $S$, $E$, $I$, and $R$.
	\item Mix the partitions by disease category: $M_S S$, $M_E E$, $M_I I$, $M_R R$.
	\item Integrate the updated S, E, I, and R vectors within each partition over as many time steps as is the mixing duration in days.
	\item Unmix the partitions by disease category: $M_{S'} S' , M_{E'} E', M_{I'} I', M_{R'} R'$
	\item Distribute the $\Delta S_n, \Delta E_n, \Delta I_n, \Delta R_n$ from integrating each partition to all partitions m that travelled to partition n for the mixing event
\end{enumerate}

\begin{figure}[htbp]
    \centering
    \includegraphics[width=0.4\linewidth, trim = 0cm 1cm 0cm 0.2cm, clip]{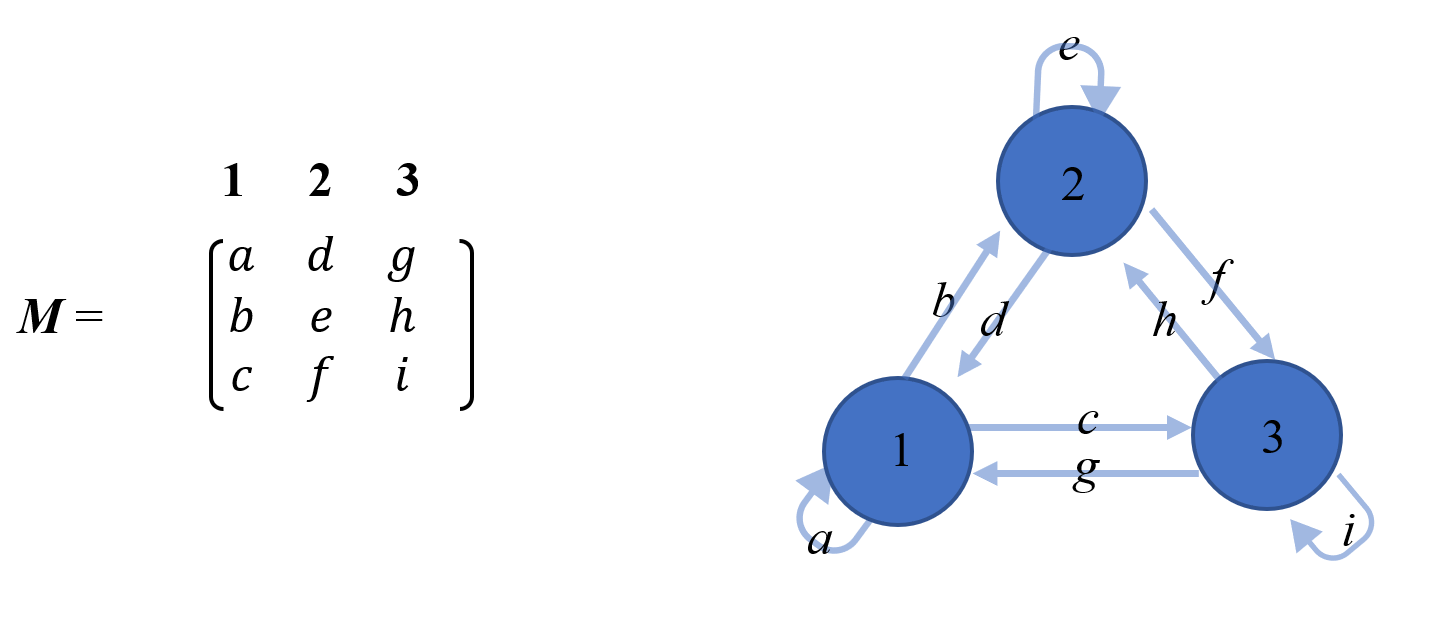}
    \caption{A mixing matrix $M$ can be thought of as the weighted, directed edge values of a graph in which each node represents a partition of the population.}
    \label{fig:mixing}
\end{figure}

\noindent \textsf{\textbf{\small{Step 1: Generating Mixing Matrices}}}

Element $M_{i,j}$ of each matrix represents the fraction of partition $j$ that travels to partition $i$ for the mixing event. Each column of matrix $M$ must sum to one (a left stochastic matrix).

Different types of matrices can represent different types of mixing events in real life. Two examples of matrices developed thus far represent gatherings at large events in a single partition and random intermixing between all partitions. While there are numerous other mixing matrices that have been and can be developed, these two are explained in greater detail below.

The large gathering matrix is generated using an inputted probability $p_\text{stay}$ that members of each partition will stay home rather than travel to the large gathering. The current model uses the same $p_\text{stay}$ for each partition, but $p_\text{stay}$ can vary to model different behavior among partitions. Members of each partition will either stay in their partition with probability $p_\text{stay}$ or travel to the large event (hosted in the last partition) with probability 1-$p_\text{stay}$. An example mixing matrix for \texttt{Num\_boxes} $= 3$ is:
\vspace{-0.2in}
\begin{center}
\[
\begin{pmatrix}
p_\text{stay} & 0 & 0\\
0 & p_\text{stay} & 0 \\
1 - p_\text{stay} & 1 - p_\text{stay} & 1
\end{pmatrix}
\]
\end{center}

The random mixing matrix is generated by summing an inputted number $m$ of randomly generated permutation matrices, each multiplied by its own randomly generated decimal between 0 and 1. The columns of the summed matrix are then normalized to ensure the matrix is left stochastic. The larger values of $m$ correspond to higher levels of mixing between partitions - the number of zero entries (meaning nobody from partition $j$ travels to partition $i$) decreases as more permutation matrixes are summed together. This is one way of controlling the amount of mixing in the model. (Note if $m$ = 1, this is the equivalent of no mixing because the partitions are simply switched around, but the groups remain isolated).

There is also a uniform mixing matrix where each partition mixes in each other partition with a uniform probability equal to a constant, population-wide probability of staying home and not mixing. Another matrix, a multi-modal travel matrix, has been developed, that is essentially a convex combination of mixing matrices with different levels of sparsity: If we consider each mixing matrix to represent one type of travel, \emph{e.g.}, short versus long distance travel, then a combination of mixing matrices can represent a combination of different modes of travel. A \emph{convex} combination simply ensures that the resulting matrix is still a mixing matrix where each column sums up to one.

\vspace{0.1cm}
\noindent \textsf{\textbf{\small{Step 2: Calculating Unmixing Matrices}}}

We developed an unmixing matrix that uses the original mixing matrix M and the initial $S$, $E$, $I$, and $R$ vectors ($v$) to ensure the same number of people who travel to partition $j$ from partition $i$ during a mixing event leave partition $j$ at the end of the mixing event to return to partition $i$.The formula to calculate each element of the unmixing matrix $I$ for mixing matrix $M$ and $S$, $E$, $I$, or $R$ vector $v$ is shown below. The percentage of people who return to partition $i$ from partition $j$, $I_{ij}$, equals the number of people who went to partition $j$ from partition $i$ ($M_{ji} v_i$) divided by the total number of people who travelled to partition $j$ ($M_j \cdot v$). If nobody travels to partition $j$ ($M_j \cdot v = 0$) during a mixing period, then nobody returns from partition $j$ to partition $i$ at the end of the mixing period ($I_{ij} = 0$).
\begin{equation}
I_{ij} = \frac{M_{ji} v_i}{M_j \cdot v} \mbox{\; if } M_j v \neq 0; \quad \mbox{ else } I_{ij} = 0.
\end{equation}

\vspace{0.1cm}
\noindent \textsf{\textbf{\small{Steps 3-6: Attributing Changes in $S$, $E$, $I$ and $R$ Groups During Mixing to the Proper Partitions}}}

Whether there is mixing or not, the equations and parameter values that are integrated in each partition over the course of the simulation remain constant unless otherwise adjusted. 

The only difference between running the simulation during a mixing event and running the simulation in normal, isolated conditions is that the changes in the total $S$, $E$, $I$, and $R$ groups over the duration of the mixing event must be manually re-distributed to each partition after unmixing. For example, say the simulation is run for 10 days and over that time, the number of susceptible individuals in partition $i$ decreased by 25. If those 10 days were not during a mixing event, then that means there are 25 fewer people in partition $i$ who remain susceptible. If those 10 days were during a mixing event, it is not necessarily those native to partition $i$ who transitioned from $S$ to $E$. Instead, each partition $j$ with members who travelled to partition i during the mixing event ($M_{i,j} \neq 0$) may have had members transition from $S$ to $E$ before returning to partition $j$ at the end of the mixing event. 

\begin{figure}
    \centering
    \includegraphics[width=1\linewidth]{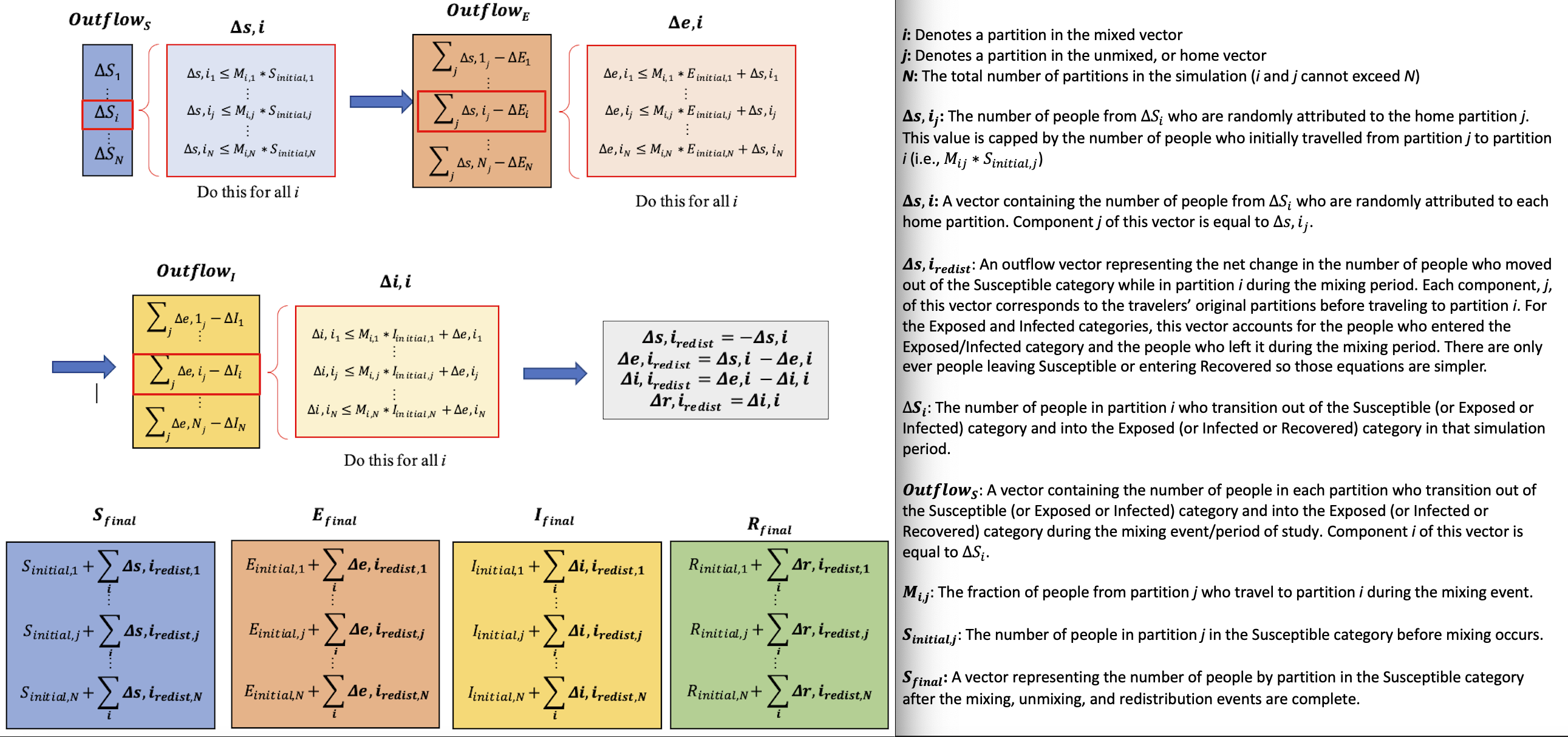}
    \caption{Each box above can be thought of as a vector. The diagram walks through the steps of reallocating each transition from one disease category to the next to the original partitions. Note that $N$ is used in place of \texttt{Num\_boxes} in this figure to represent the total number of partitions more compactly.}
    \label{fig:unmixing}
\end{figure}

Because this is an equation-based model and does not track individuals, it is impossible to know exactly which members from which partitions transitioned from $S$ to $E$ or $E$ to $I$ or $I$ to $R$ over the course of a mixing event. To account for this, the changes in $S$, $E$, $I$, and $R$ in partition $i$ are randomly distributed across all partitions $j$ in units of 1 (to represent one person), with the change in partition j bounded by the total number of people that could have transitioned from one state to another from partition $j$. This procedure of determining the bounds is shown in Figure \ref{fig:unmixing}. 

\subsection{Parameter Values and Mixing Duration}

Over the course of the mixing duration, the parameter values remain constant within each partition $i$ regardless of whom from partition $j$ has travelled to partition $i$ for the mixing event. However, there is an option to update each partition’s parameter values during a mixing event to model any changes in behavior or policies related to the mixing.

It is important to note that in many ways unmixing at the end of the mixing duration can simply be thought of as a second mixing event. The equation-based nature of this model means there is no difference between each individual of the same compartment ($S$, $E$, $I$, or $R$). So even though the population is unmixed to return everyone to the home partitions, this is essentially another mixing event as the equation-based model sees no difference between members in their home partitions and members visiting other partitions.

\subsection{Assumptions and Limitations}

This model works best to study the effects of sporadic, infrequent mixing events rather than situations with continuous mixing between partitions. This means it is best used to study what happens when populations that naturally have separation between groups whether due to geography or pandemic policy come into occasional contact. Complete separation of partitions at all times other than during the mixing event is an unrealistic expectation, but the model uses that assumption to balance complexity and robustness. Rather than focusing on the impact of mild, regular mixing events such as trips to the grocery store, this model helps to understand the effects of intense, irregular mixing events such as university students travelling during breaks or  large sports events. Results produced by this model may be an underestimate of the disease outbreak severity as it is not accounting for mild, regular mixing. As such, it is best used to understand the relative impacts of different variables on outbreak severity rather than as an absolute measure of how many people will get infected.

The model assumes homogeneous mixing within partitions which may not always be the case. When partitioning a population to study in this model, it is best to choose partitions that divide people into group sizes roughly matching the scale at which homogeneous mixing takes place. In other words, treating countries as partitions in this model (while also computationally very demanding due to the large population sizes) may be less indicative of real trends than using counties as partitions due to the widely varying behavior of individuals across entire countries. Just as too large of partitions may affect the model's accuracy, if the partition size is too small, the results will likely be unstable as the sample size is too small to reliably predict average behavior. The model itself will run successfully for any partition size greater than zero, and the upper bound on partition size is limited by the computational power of the computer. 

The equation-based nature of this model means rather than using whole numbers to properly represent people, the evolution of populations in the $S$, $E$, $I$, and $R$ groups within each population are rational numbers. These fractions of people in each group which are not possible in reality can enable second waves of outbreaks that would not have otherwise happened – small decimals of infected people enable infections to persist longer than in real life. Furthermore, the matrix multiplication for mixing also allows fractions of people to travel to other partitions. While this model does use whole numbers for redistributing individuals after mixing events, the other times in which fractional people are allowed may accumulate and cause results to vary from reality.

\section{ANALYSIS}
\label{sec:analysis}


\begin{definition}[Trip, Travel, Travel Cadence]
A \emph{trip} is an origin-destination pair. 
A \textit{travel} period consists of multiple trips. \emph{Travel cadence} is a tuple, (number of trips, total travel duration), that characterizes the travel period. For example, the two trip, 10 day travel period tuple (2, 10) represents people traveling from one location to another for 5 days. After 5 days, they travel to another location and stay there for the remaining 5 days of the travel period. (2,10) is the travel cadence.
\end{definition}

To facilitate analysis, we assume that trips are synchronized for all partitions. But asynchronous trips can be implemented with our model as well (albeit with more coding effort - numerous mixing matrices and mixing events are needed to allow specific partitions to mix with other partitions at distinct times).

\begin{definition} [Epidemic Spread]
We define the \emph{epidemic spread} (for a fixed period and a particular population) as the number of infections occurred during that period.
\end{definition}

The main technical question we want to answer is: \emph{What is the impact of travel cadence on epidemic spread}? In other words, we would like to know the mapping between cadence and spread.
To this end, we design a sequence of numerical experiments to examine how the two dimensions of travel cadence impact the epidemic spread.

\subsection{Experiment Setup}

In the first set of numerical experiments, we focus on simulating population mixing in the origin and destination partitions. That allows us to single out the effect of interactions between travelers and residents in every partition. For the second set of numerical experiments, we add an additional step in the simulation to also include the potential interactions among travelers themselves, who may have different origin-destinations. This second type of mixing can be attributed to, for example, interactions at airports, train stations, and bus terminals.

The experiments are conducted by running a unique simulation of our epidemic model for each travel cadence tuple under test. This corresponds to 36 data points per experimental trial (combinations of zero to three trips and 2, 4, 8, 10, 14, 20, 40, 60, or 80-day travel periods). Each simulation begins with a 30-day pre-mixing period to allow time for the infection to spread. The next 80 days of the simulation vary in amount of time spent mixing or in home partitions based on the travel period and number of trips. There are a final 14 days of the simulation without any mixing to allow infections from the end of the mixing period to develop and recover fully before measuring the total spread. Note that all tuples with zero trips are identical because there is no travel in this case so individuals remain in their home partitions for the entire trip. The zero trip trials serve as the “constants” or reference cases in the experiment. 

At the end of each simulation, we calculate the epidemic spread as the total percent of the population in the ``Recovered'' group on day 124 less the total percent of the population in the ``Recovered'' group on day 30, immediately before mixing begins. This allows for a comparison of the effects of each travel cadence on the number of people who become infected during the travel period. The zero trip case serves as the control of this experiment to show how the epidemic evolves without any mixing between partitions. 


For parameter calibration, we tried different numerical setups, and the results do not differ qualitatively from the current ones. We therefore only report the results obtained from a particular set of parameter values, as described below.
We chose a basic reproduction number ($R_0$) of roughly 2.4 to match the observed $R_0$ value during early stages of COVID-19 \shortcite{You2020}. More specifically, we have $\sigma = 4$ days, $\gamma = 10$ days, $p = 0.02$, $p_2 = 0.01$, $\lambda = 10$ people per day.
In other words, we assume that on each day, every individual in state $E$ would transition to state $I$ with probability $1/4$, and every individual in state $I$ would transition into state $R$ with probability $1/10$. The probability of infection when a state $S$ individual comes in contact with a state $E$ individual is 0.01, and the probability of infection when a state $S$ individual comes in contact with a state $I$ individual is 0.02. Everyone meets 10 people daily.

We set up other parameters as follows. We assume 100,000 people are distributed equally among 200 partitions. The total numbers of state $E$ and state $I$ people are 5 and 50 respectively at day 0. 
We calculate the mixing matrix as the sum of several matrices, each one representing a specific travel distance and percentage of population. More specifically, we let $M = M_1 + M_2 + M_3$, where $M_1$ takes 1\% of the population in each partition and distribute them equally in all other partitions, $M_2$ takes 5\% of the population in each partition and distributes them equally to 10\% of other partitions, and $M_3$ takes 10\% of the population in each partition and distributes them equally to 1\% of other partitions (we round up the number of receiving partitions to an integer in each case).
We vary overall quarantine rate (assumed to be the same across all partitions) and travel cadence in our numerical analysis: quarantine rate $q \in \{0, 0.1, 0.2, 0.3, 0.4, 0.5\}$, and travel cadence $\in \{0, 1, 2, 3\} \mbox{ trips } \times \{1, 2, 4, 8, 10, 14, 20, 40, 60, 80\} \mbox{ days}$.

\subsection{Numerical Results and Policy Implications}

\begin{figure}[ht!]
\vspace{-0.8cm}
    \includegraphics[width=.39\textwidth, trim = 0.2cm 5cm 0.3cm 4cm, clip]{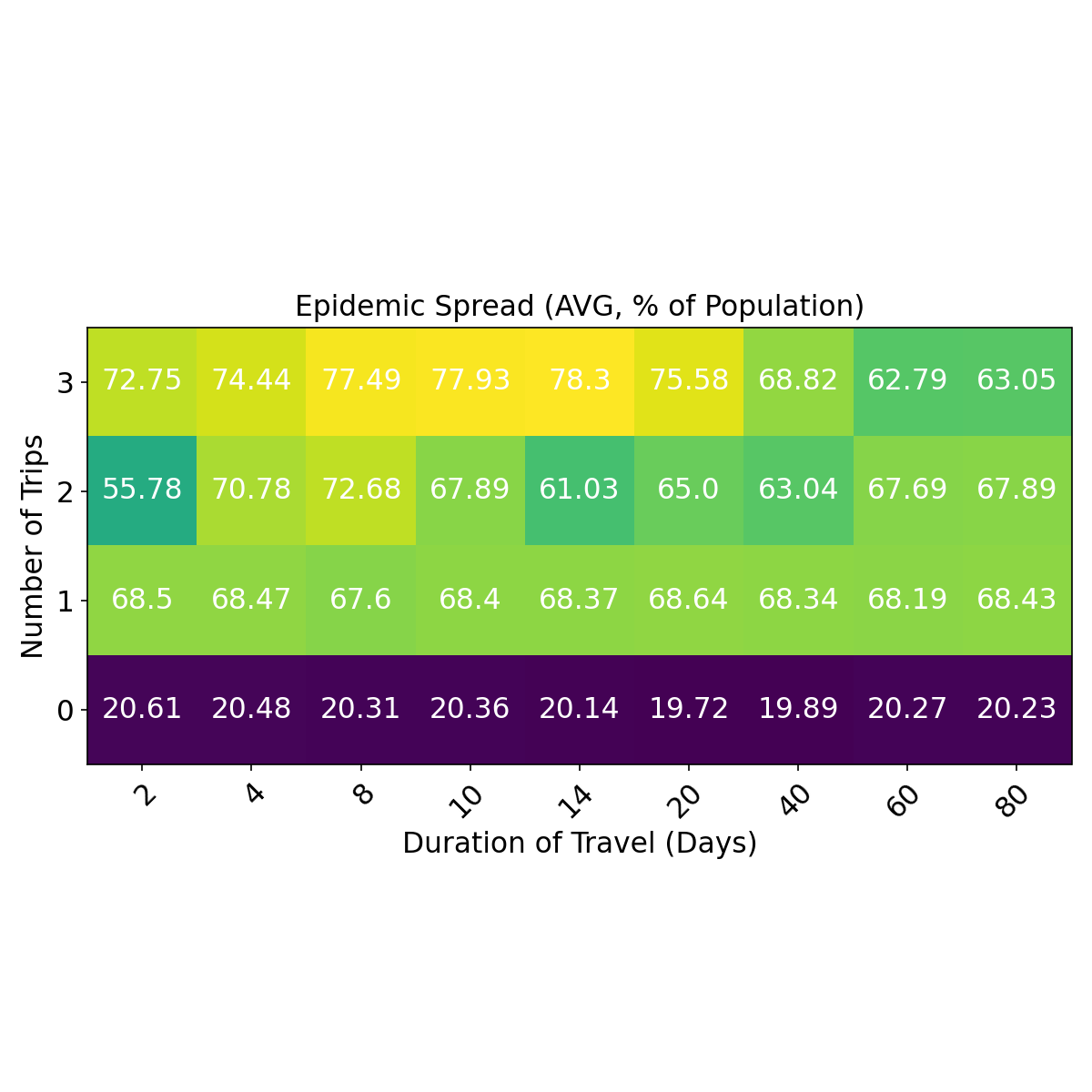}\hfill
    \includegraphics[width=.38\textwidth, trim = 1cm 5cm 0.3cm 4cm, clip]{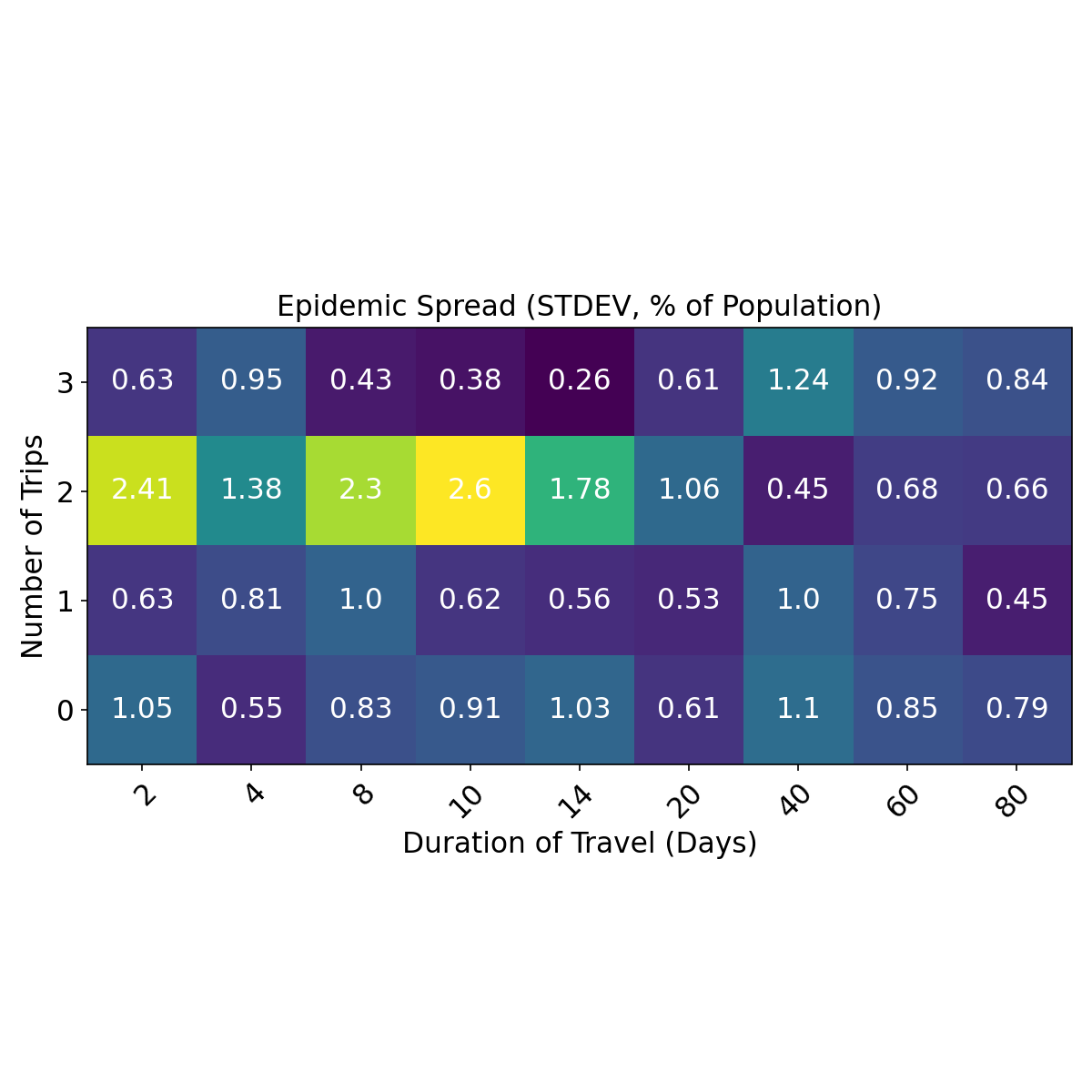}
    \\[\smallskipamount]
    \includegraphics[width=.39\textwidth, trim = 0.2cm 5cm 0.3cm 6cm, clip]{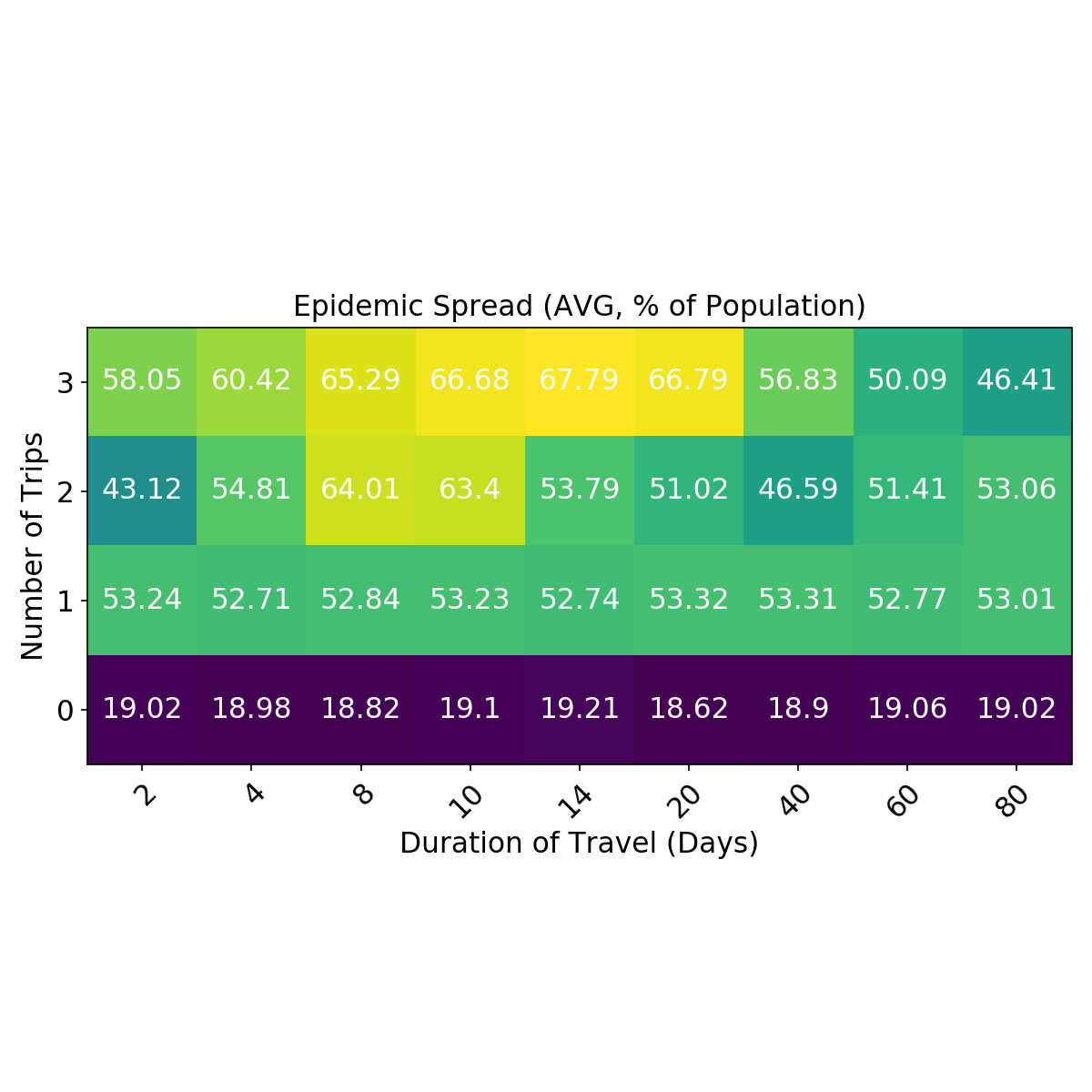}\hfill
    \includegraphics[width=.38\textwidth, trim = 1cm 5cm 0.3cm 6cm, clip]{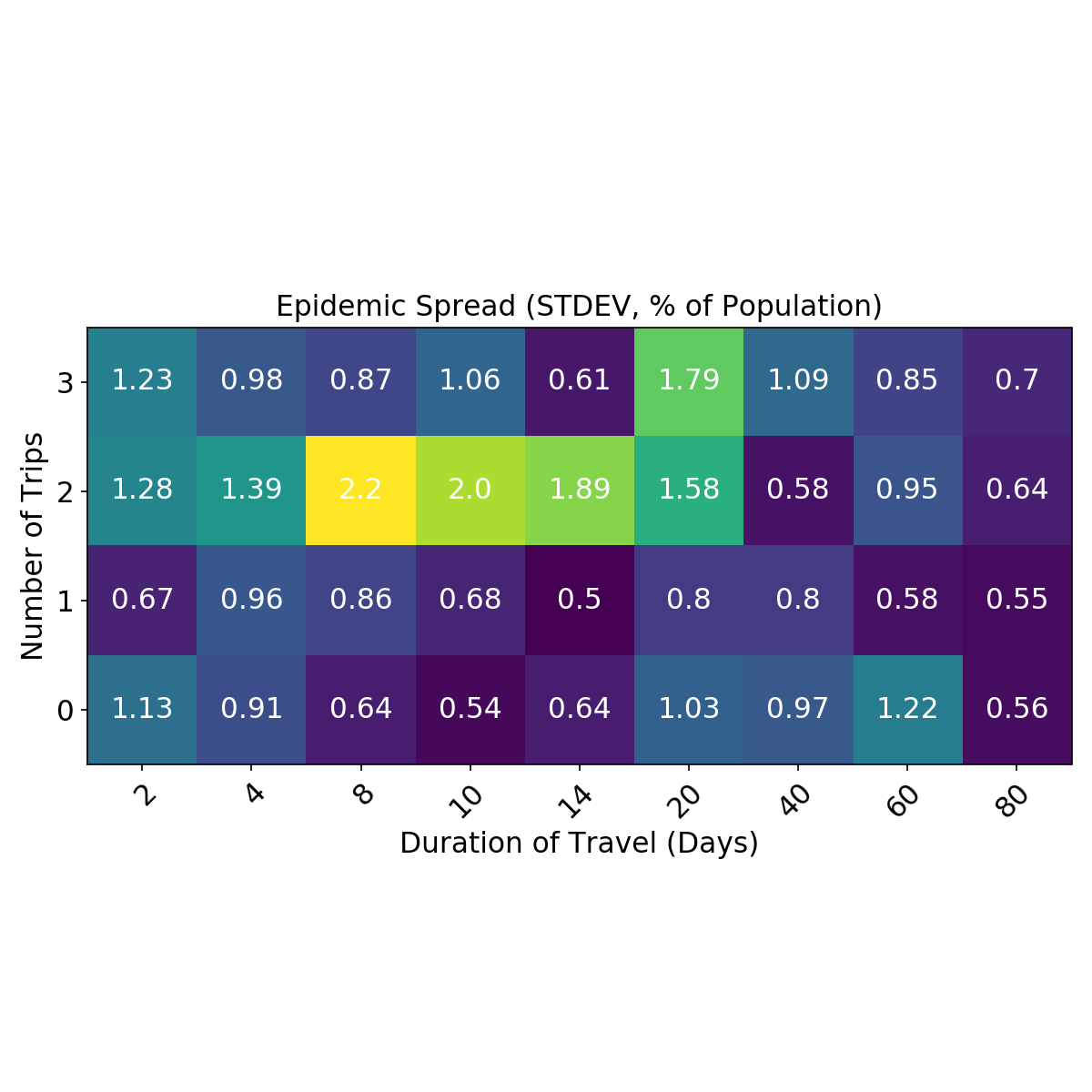}
    \\[\smallskipamount]
    \includegraphics[width=.39\textwidth, trim = 0.2cm 5cm 0.3cm 6cm, clip]{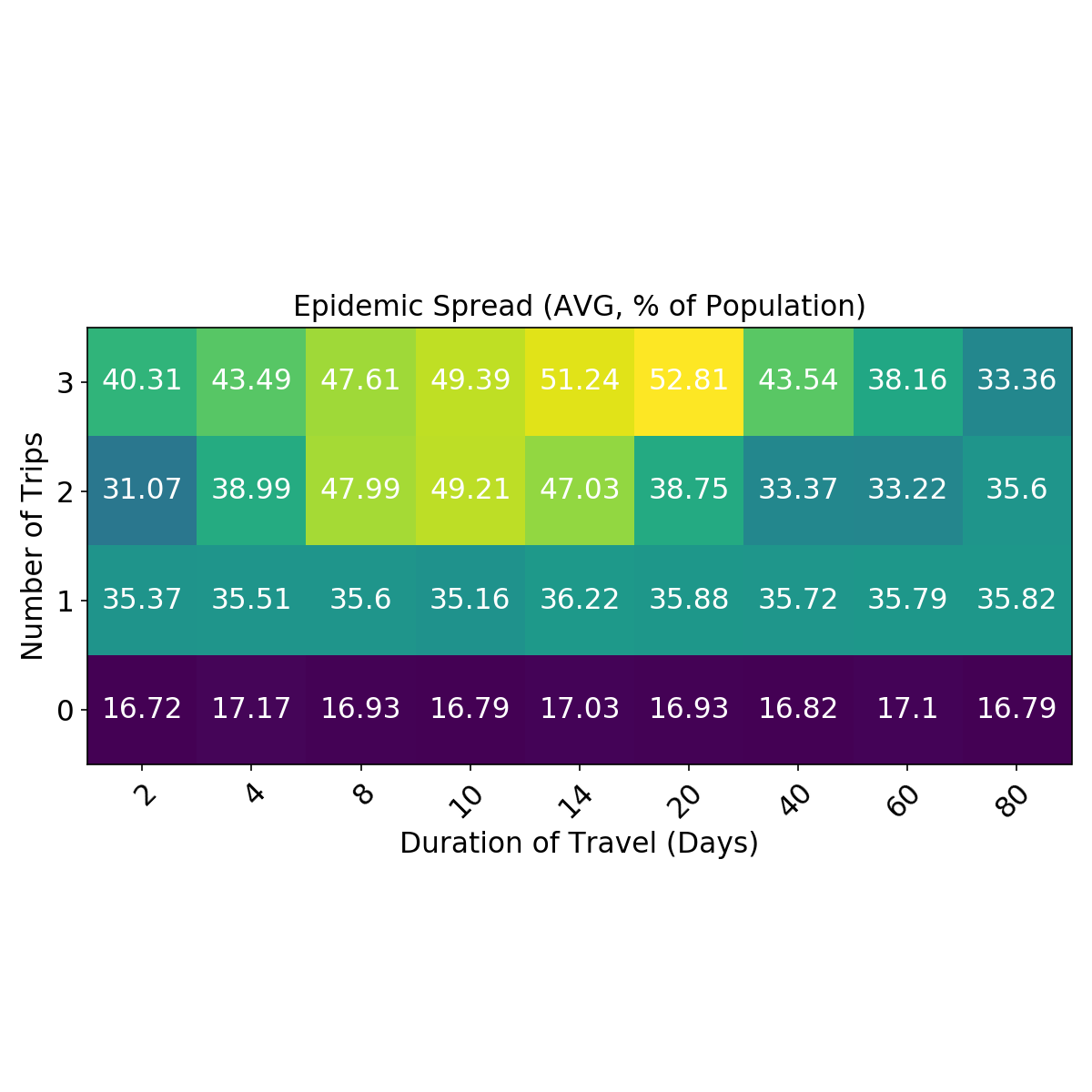}\hfill
    \includegraphics[width=.38\textwidth, trim = 1cm 5cm 0.3cm 6cm, clip]{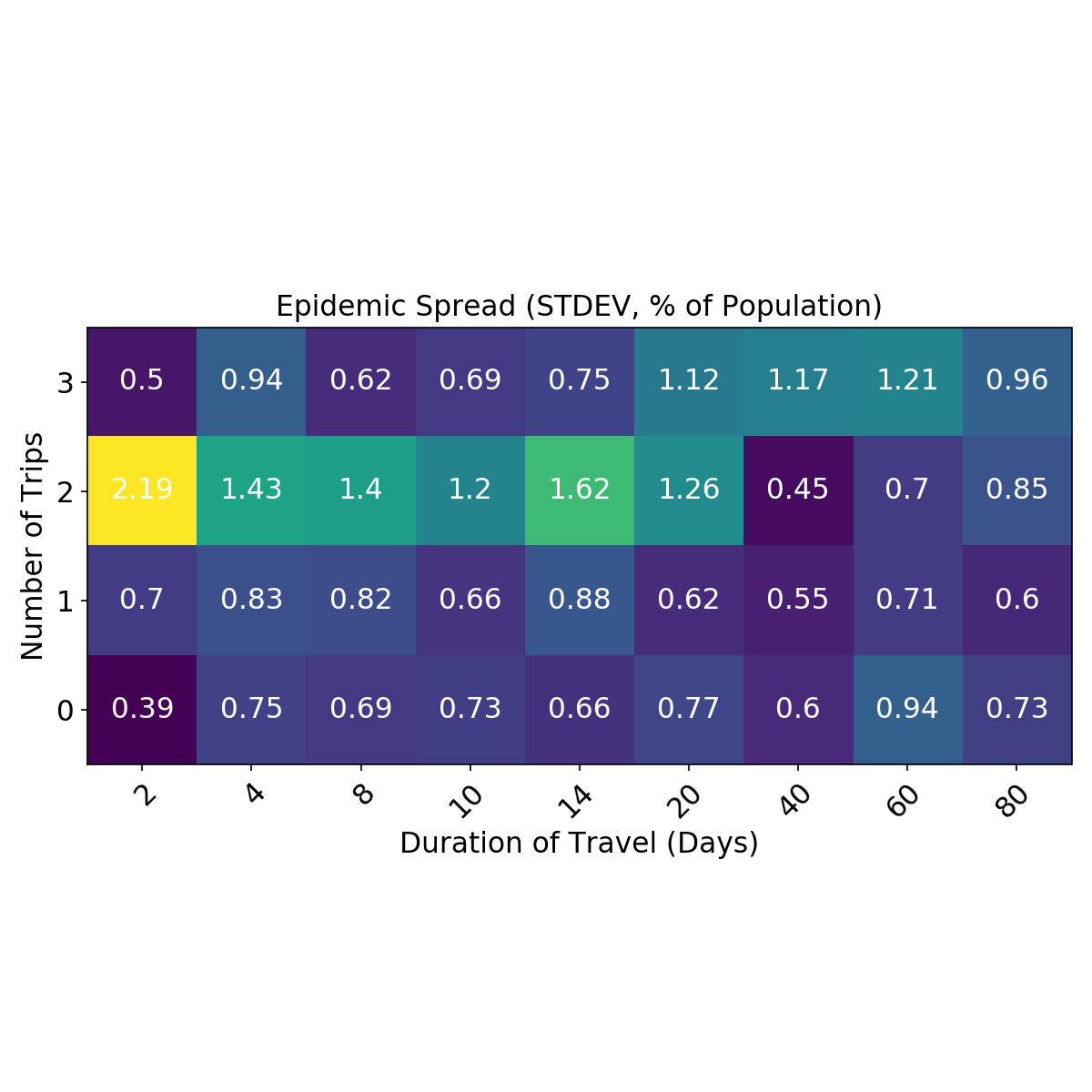}
    \\[\smallskipamount]
    \includegraphics[width=.39\textwidth, trim = 0.2cm 5cm 0.3cm 6cm, clip]{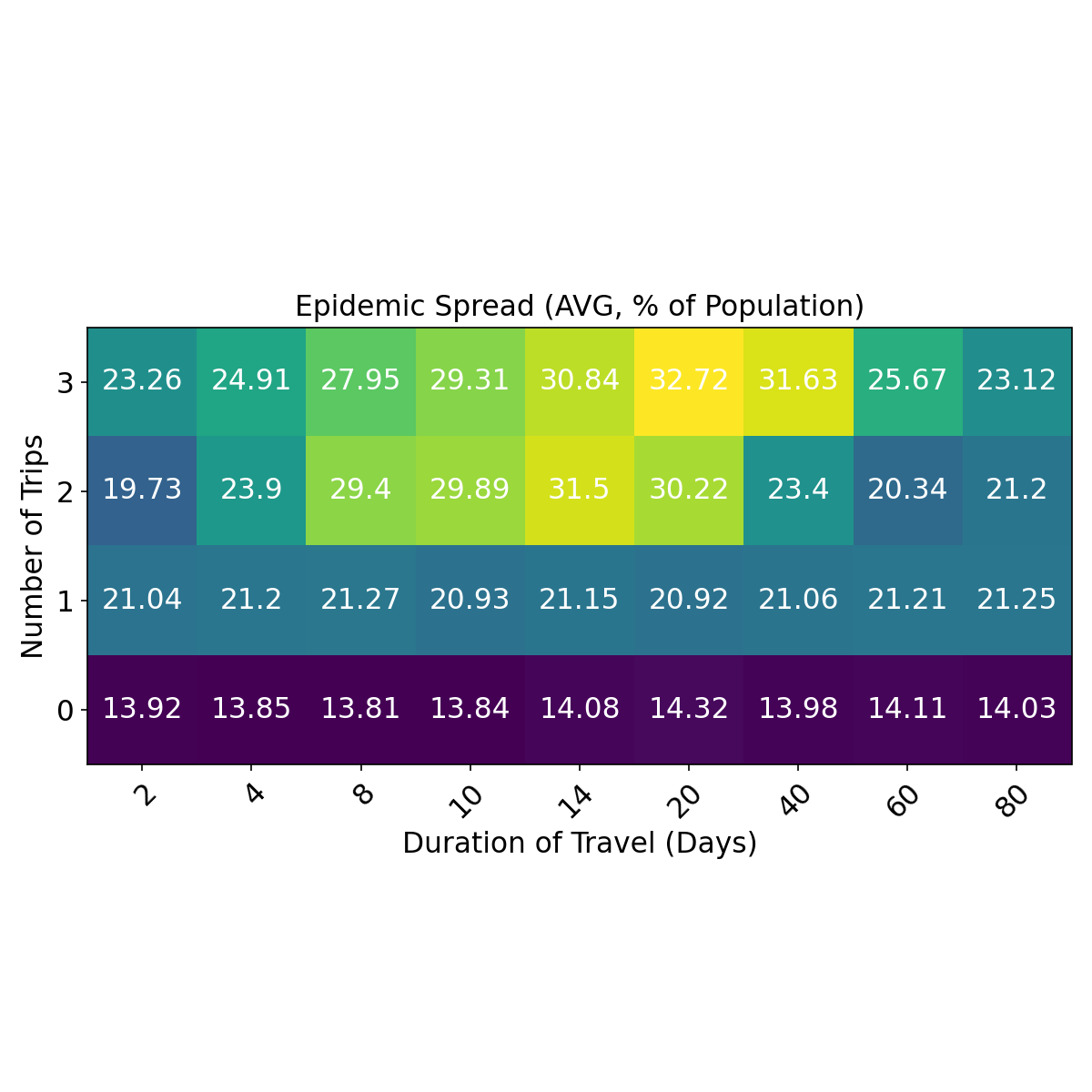}\hfill
    \includegraphics[width=.38\textwidth, trim = 1cm 5cm 0.3cm 6cm, clip]{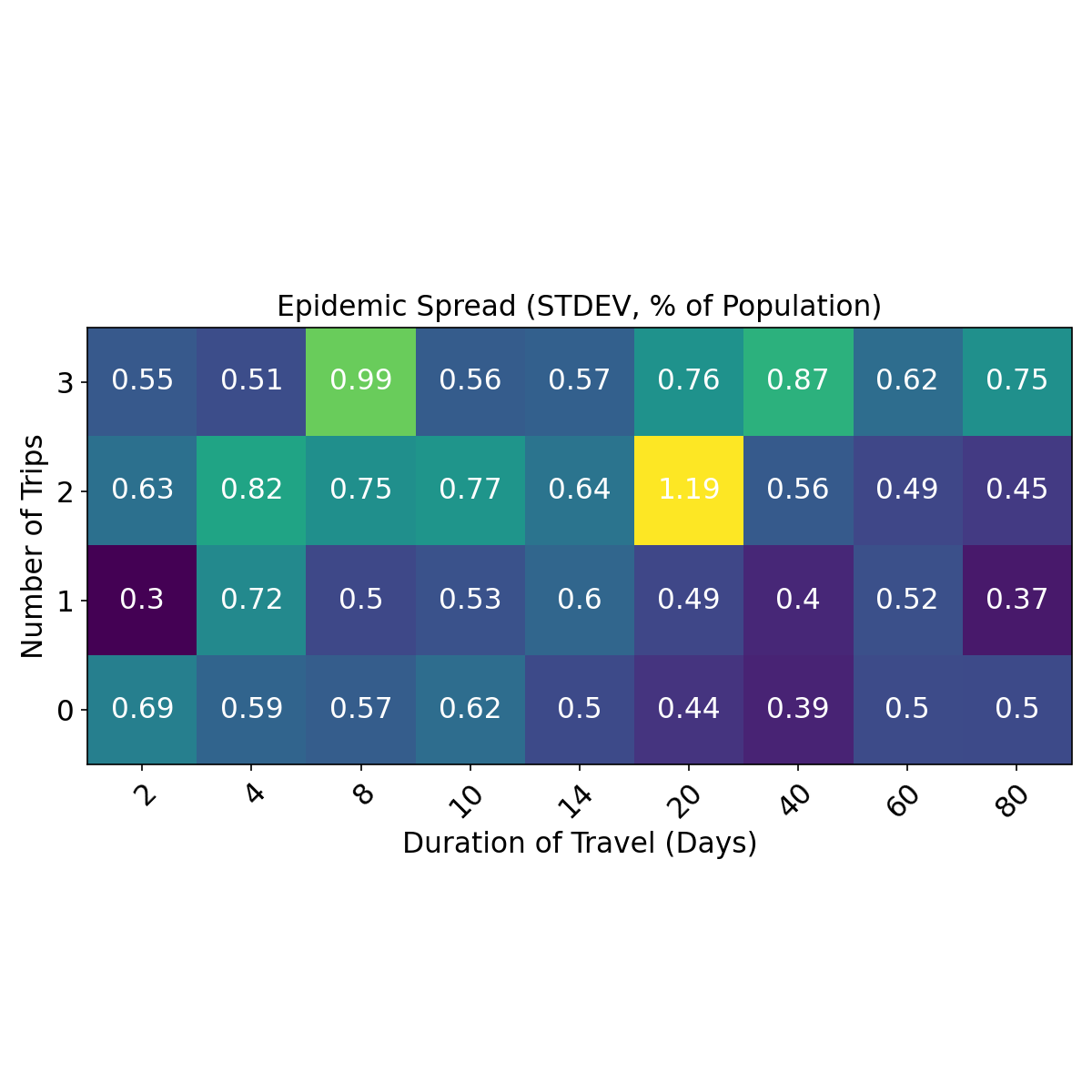}
    \\[\smallskipamount]
    \includegraphics[width=.39\textwidth, trim = 0.2cm 5cm 0.3cm 6cm, clip]{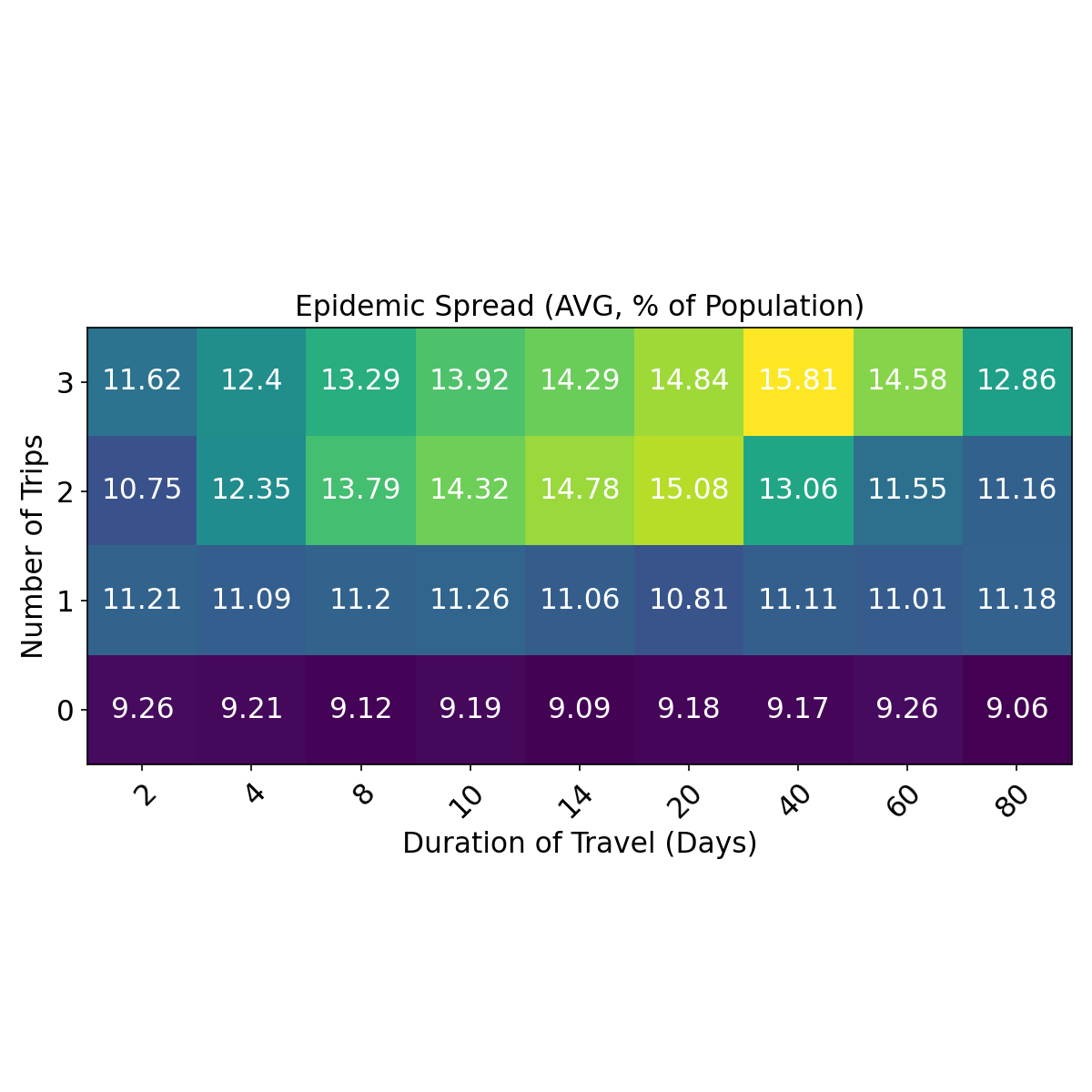}\hfill
    \includegraphics[width=.38\textwidth, trim = 1cm 5cm 0.3cm 6cm, clip]{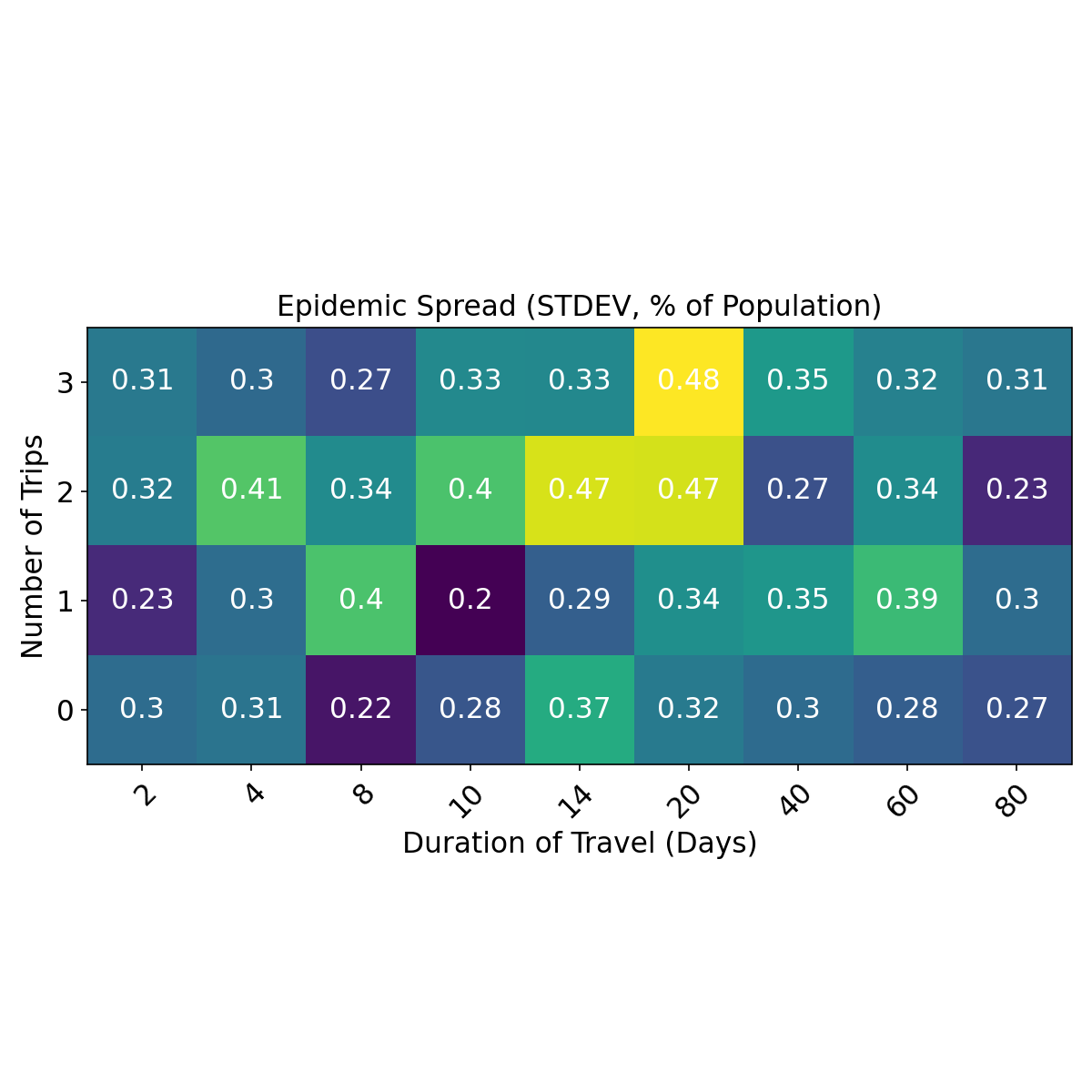}
    \\[\smallskipamount]
    \includegraphics[width=.39\textwidth, trim = 0.2cm 3cm 0.3cm 6cm, clip]{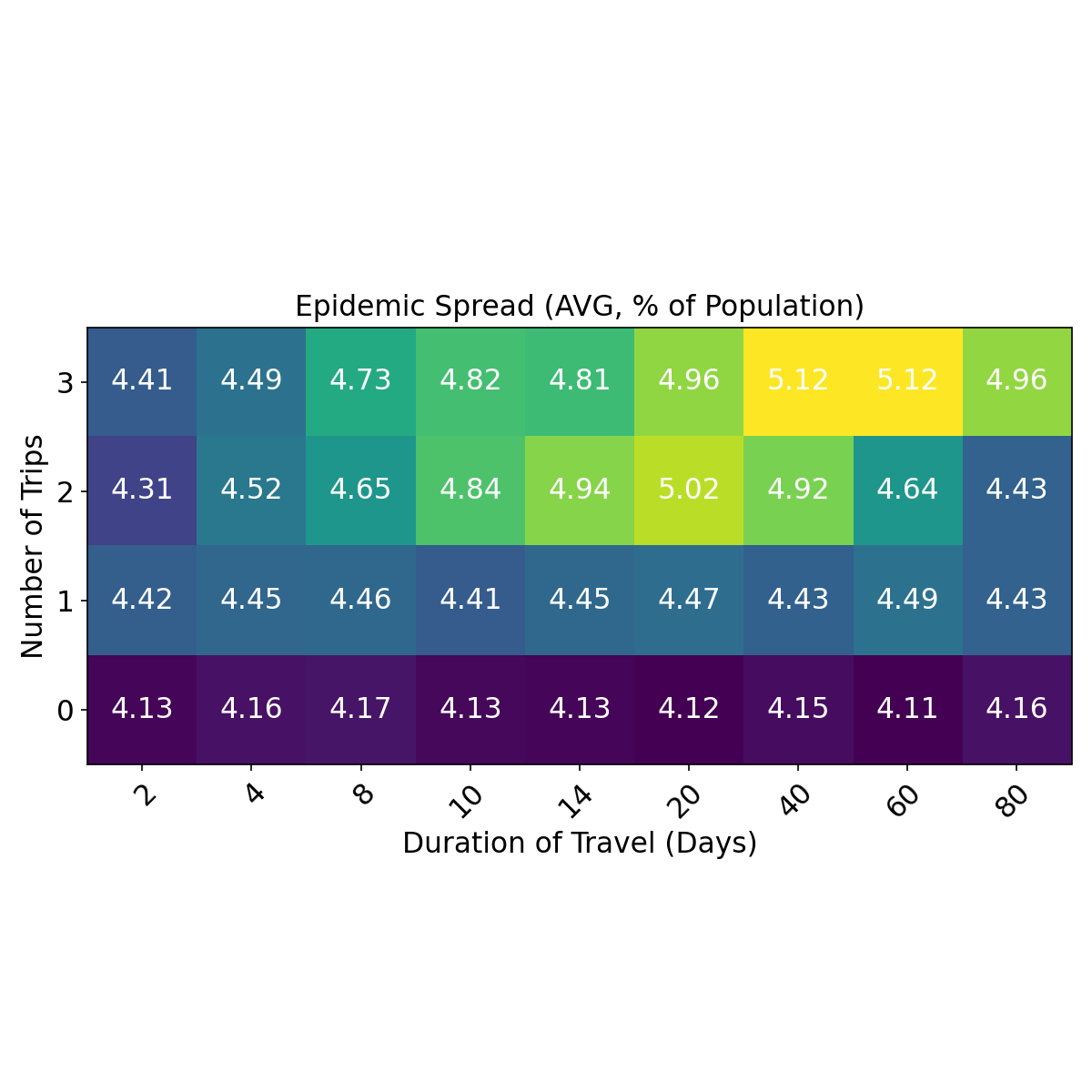}\hfill
    \includegraphics[width=.38\textwidth, trim = 1cm 3cm 0.3cm 6cm, clip]{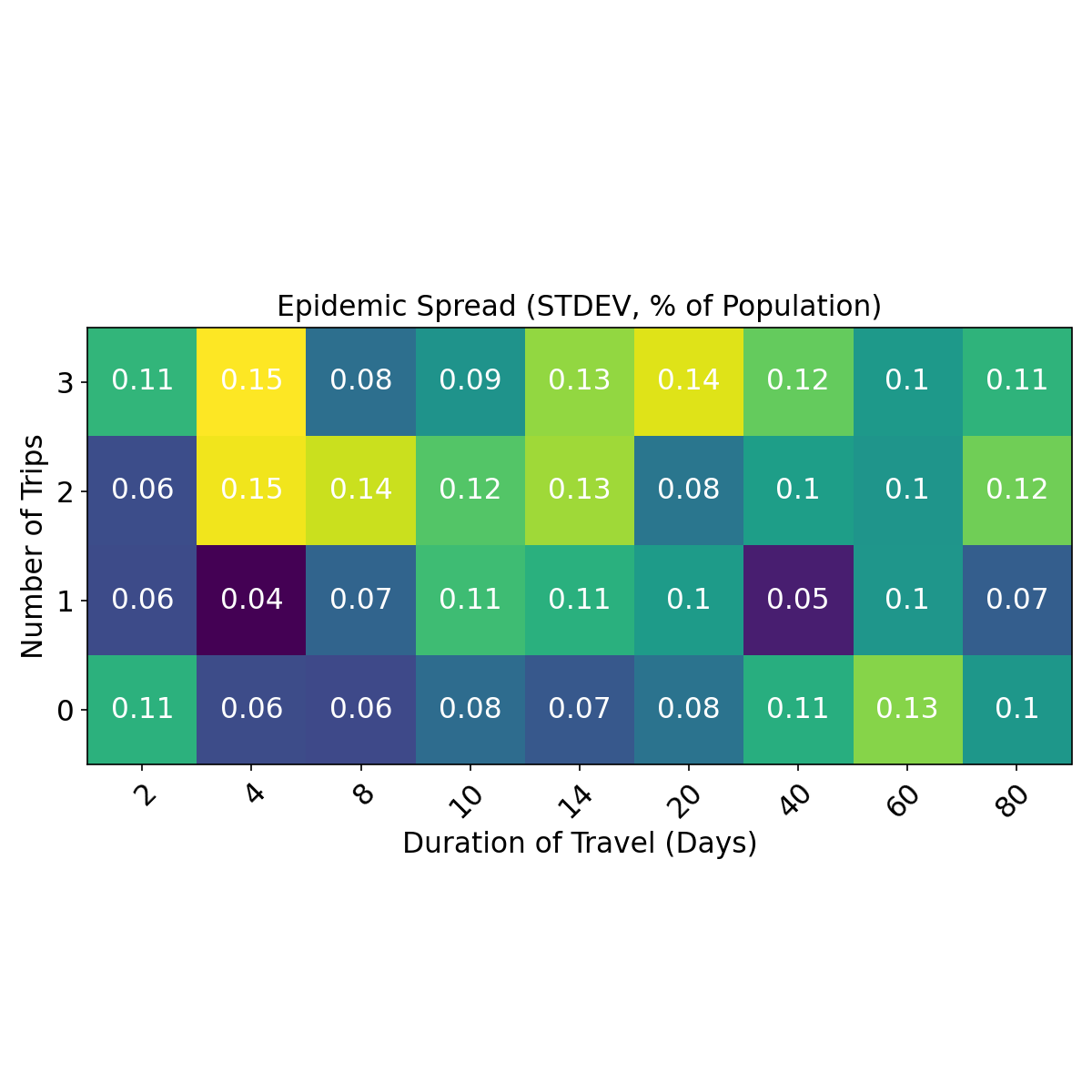}
    \caption{Epidemic spread (number in each cell) as a function of travel cadence (horizontal and vertical axes). Each row is for a different quarantine level $\in \{0, 10, 20, 30, 40, 50\}\%$. First column shows the average spread over 10 runs. Second column shows the standard deviation of spread over 10 runs.}
    \label{fig:1}
\end{figure}

\begin{figure}[ht!]
\vspace{-0.8cm}
    \includegraphics[width=.39\textwidth, trim = 0.2cm 5cm 0.3cm 4cm, clip]{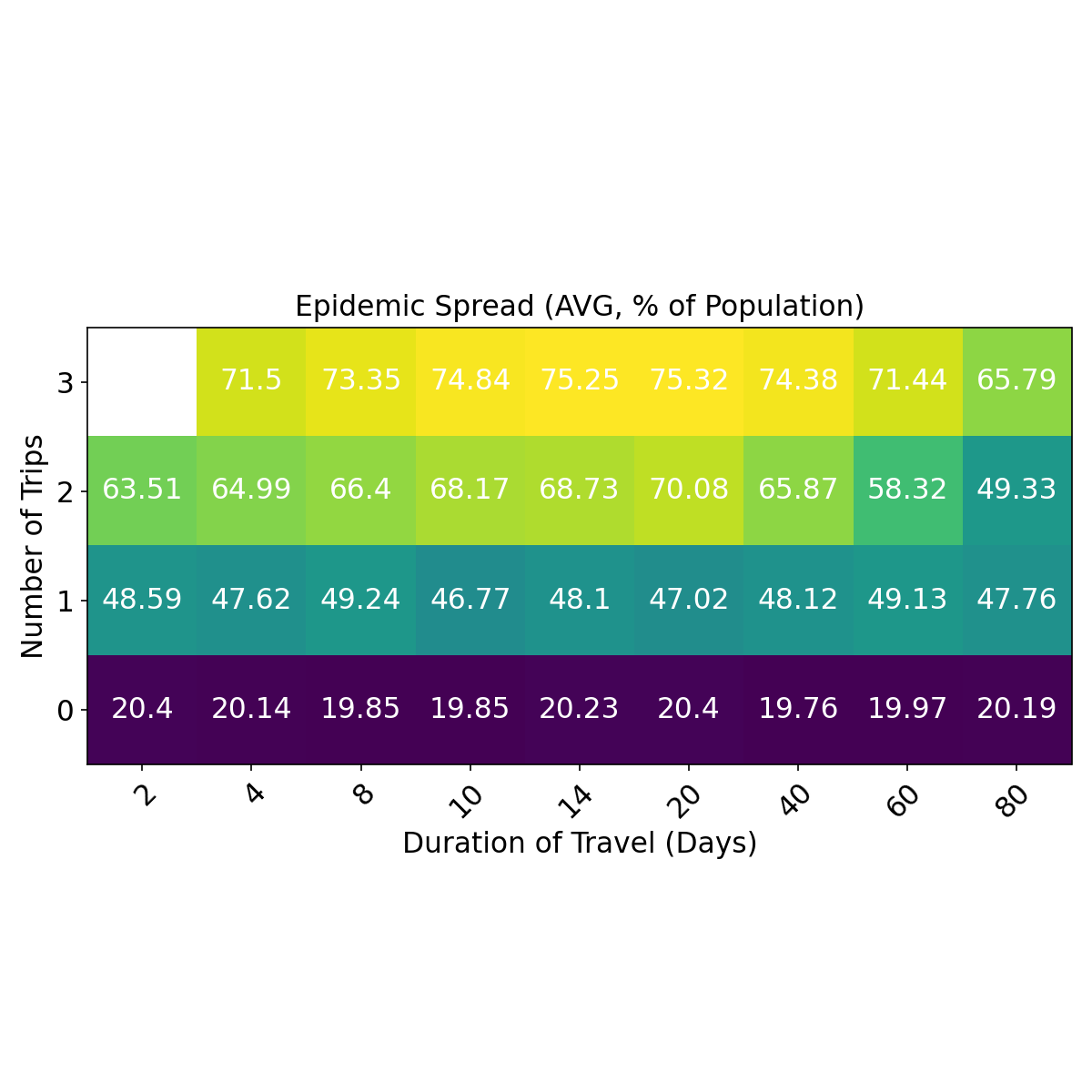}\hfill
    \includegraphics[width=.38\textwidth, trim = 1cm 5cm 0.3cm 4cm, clip]{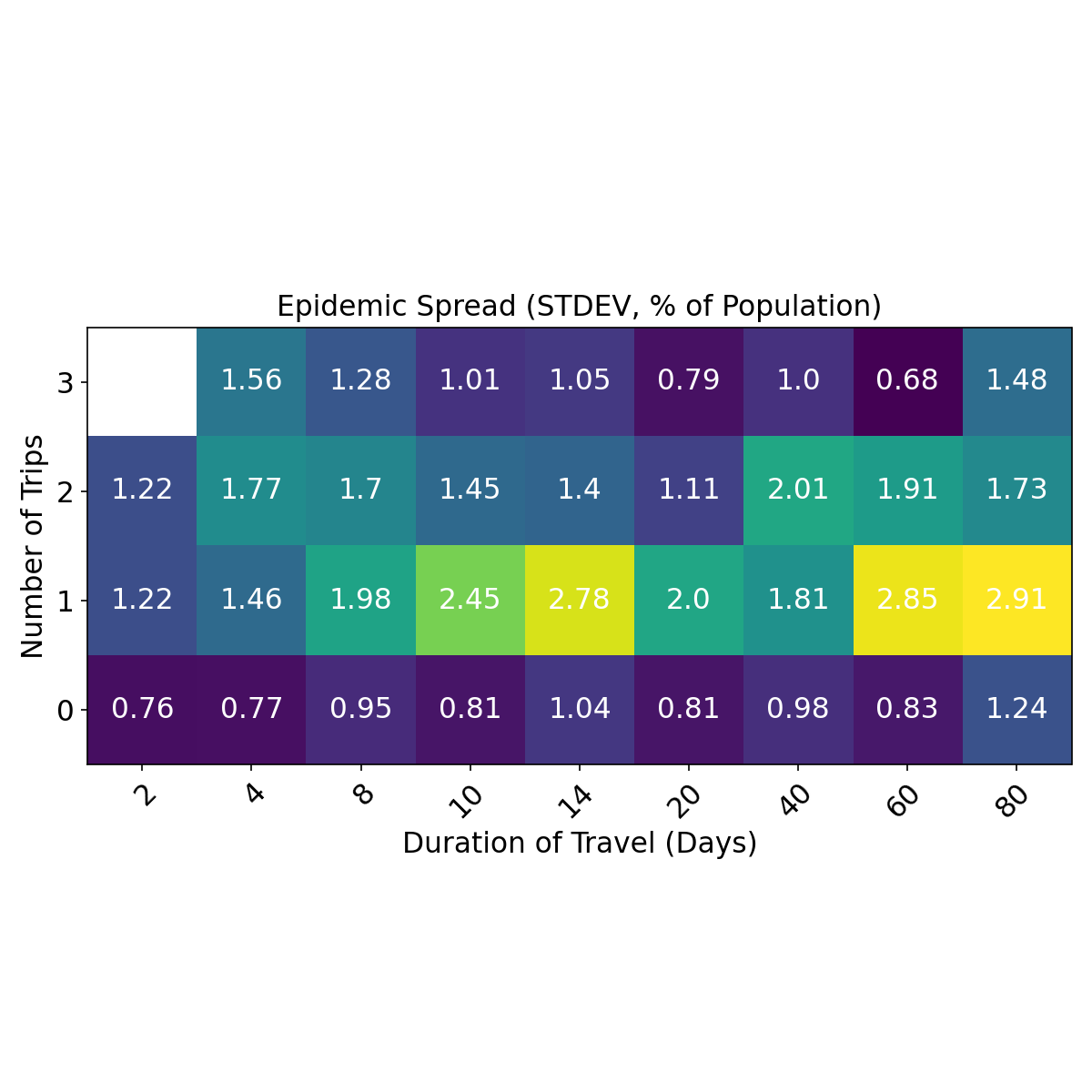}
    \\[\smallskipamount]
    \includegraphics[width=.39\textwidth, trim = 0.2cm 5cm 0.3cm 6cm, clip]{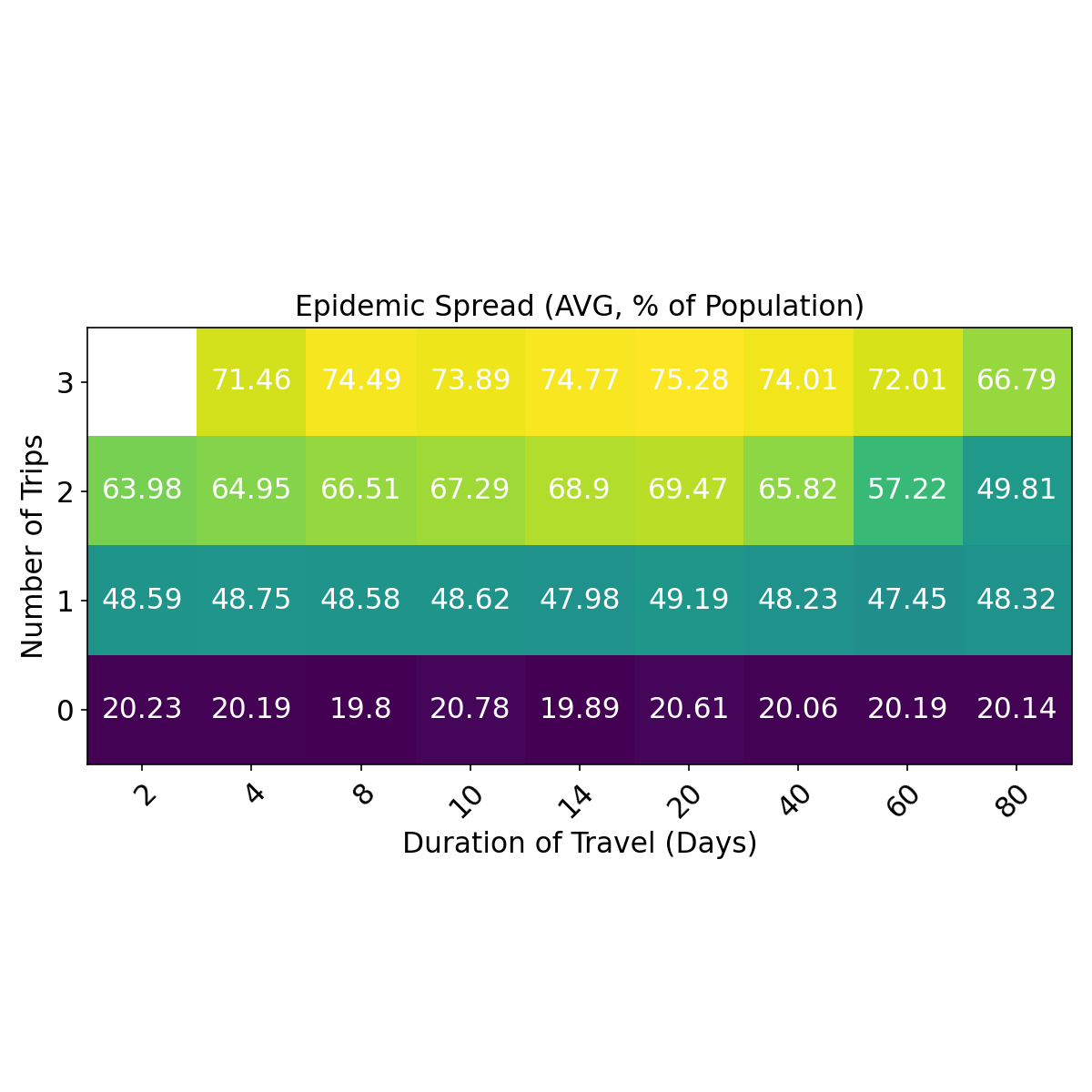}\hfill
    \includegraphics[width=.38\textwidth, trim = 1cm 5cm 0.3cm 6cm, clip]{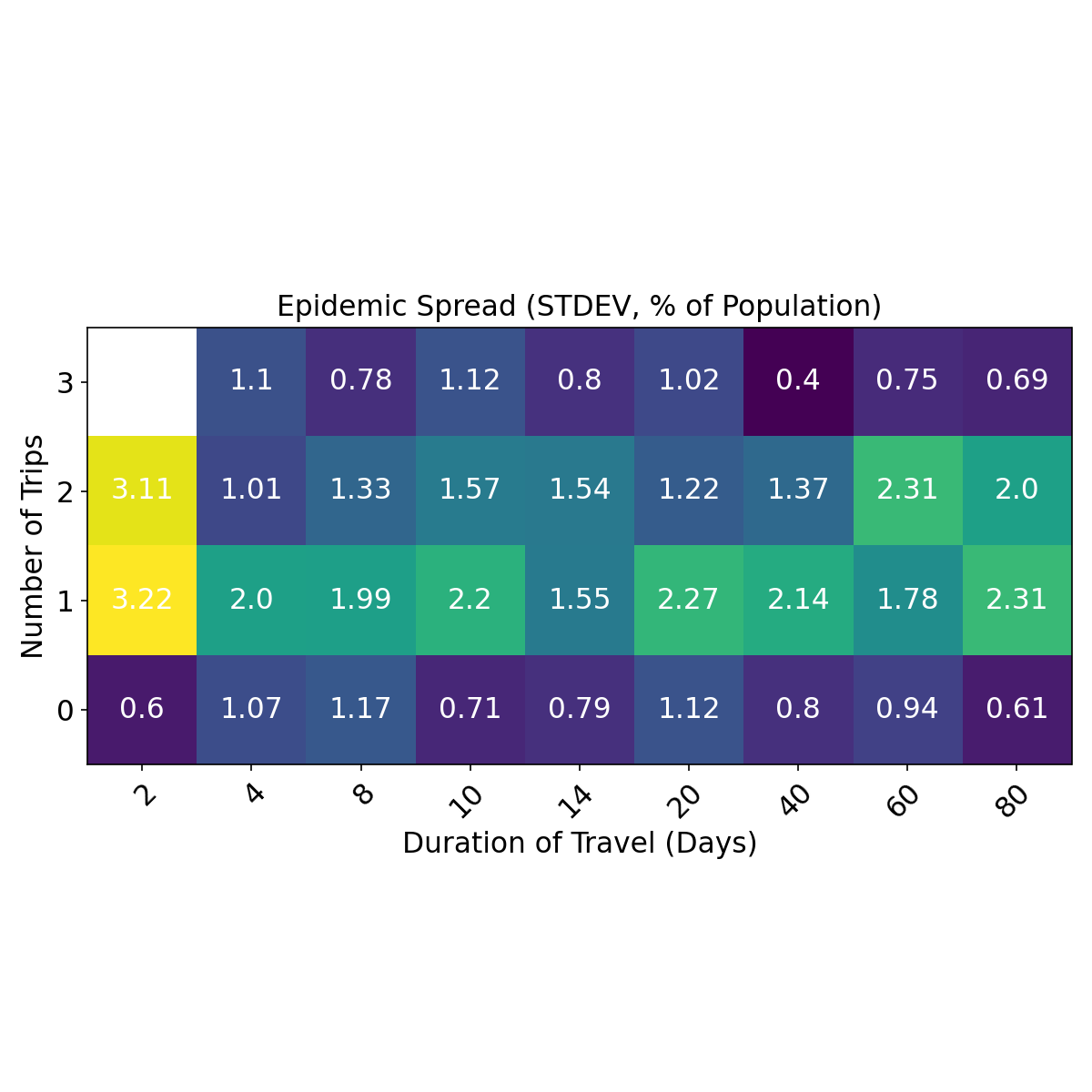}
    \\[\smallskipamount]
    \includegraphics[width=.39\textwidth, trim = 0.2cm 5cm 0.3cm 6cm, clip]{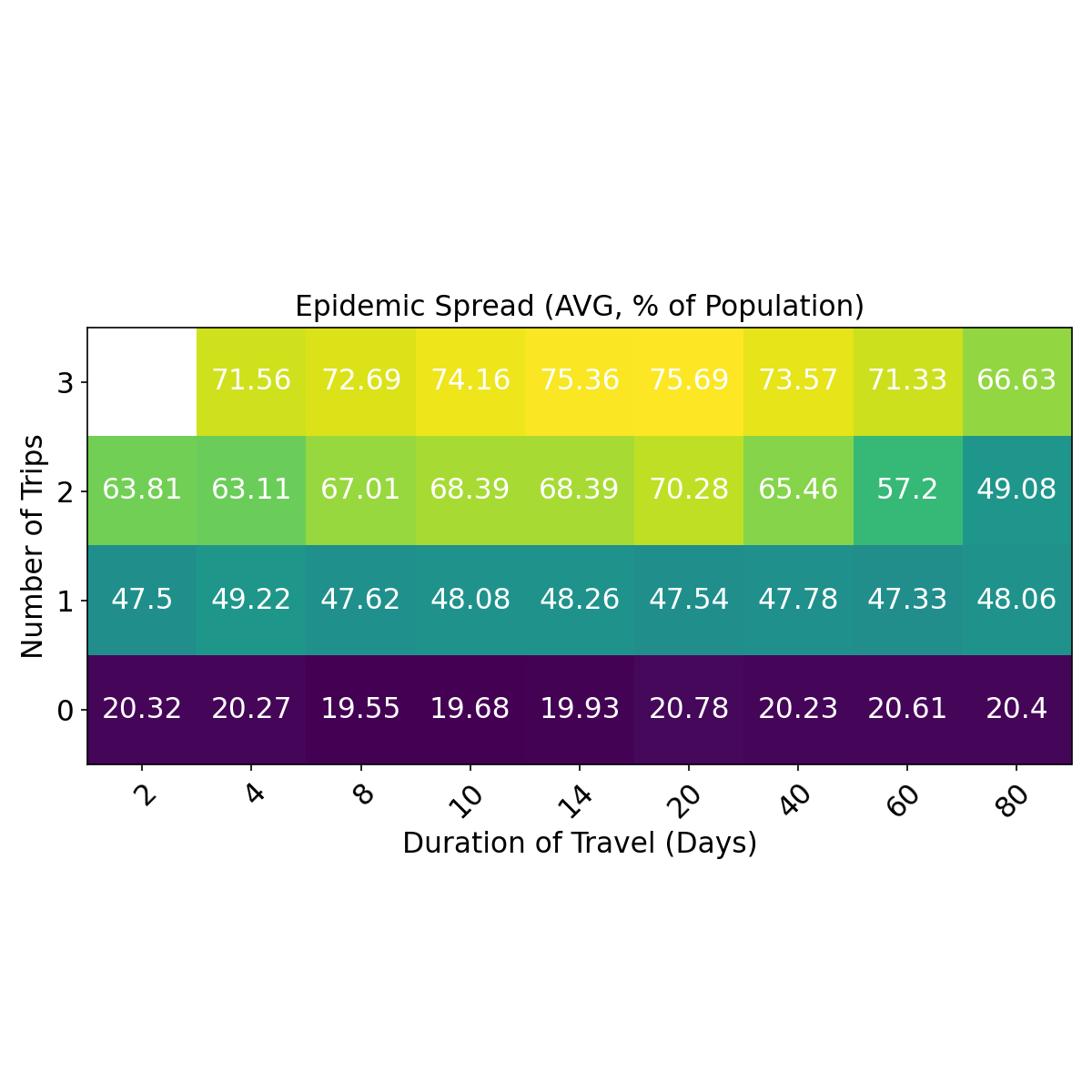}\hfill
    \includegraphics[width=.38\textwidth, trim = 1cm 5cm 0.3cm 6cm, clip]{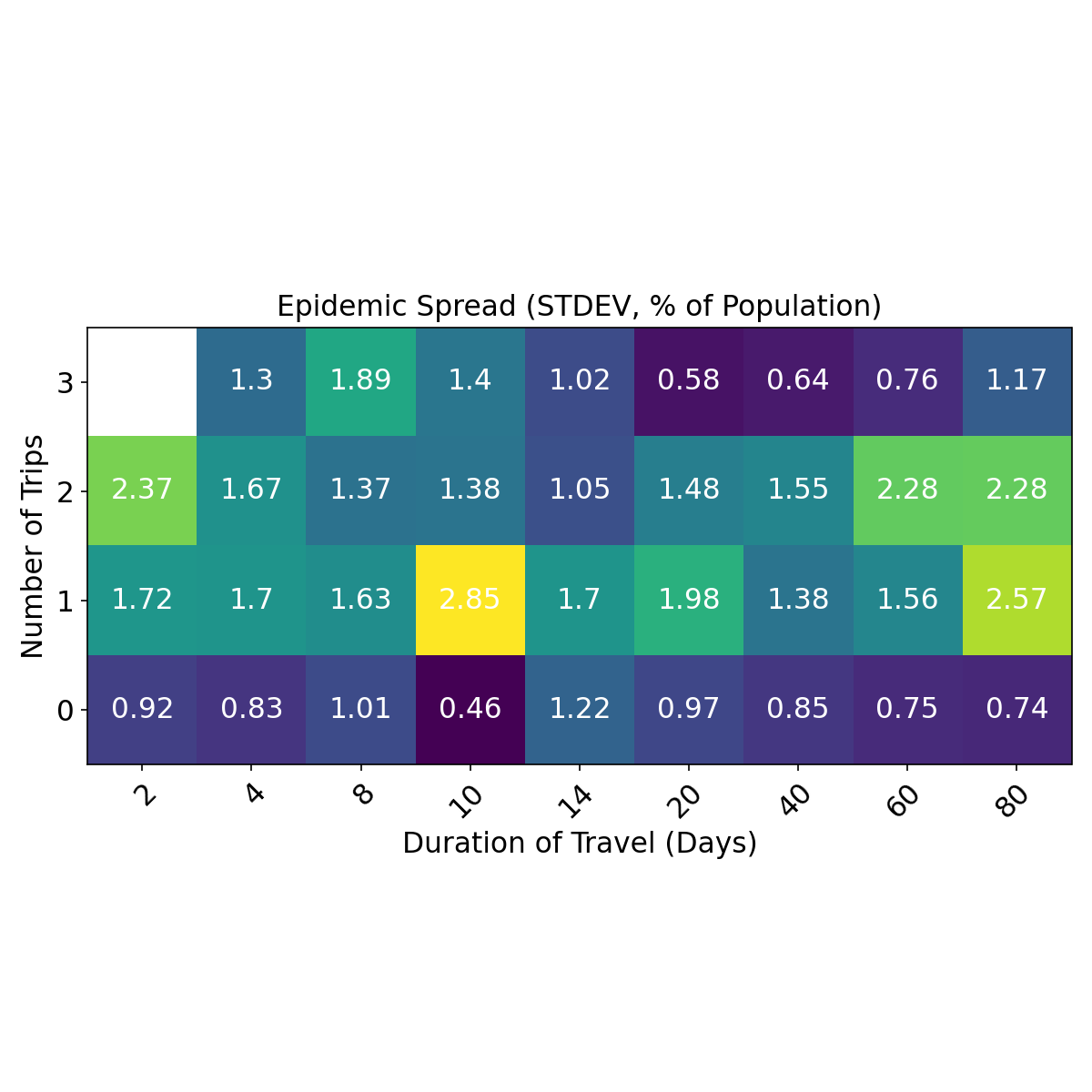}
    \\[\smallskipamount]
    \includegraphics[width=.39\textwidth, trim = 0.2cm 5cm 0.3cm 6cm, clip]{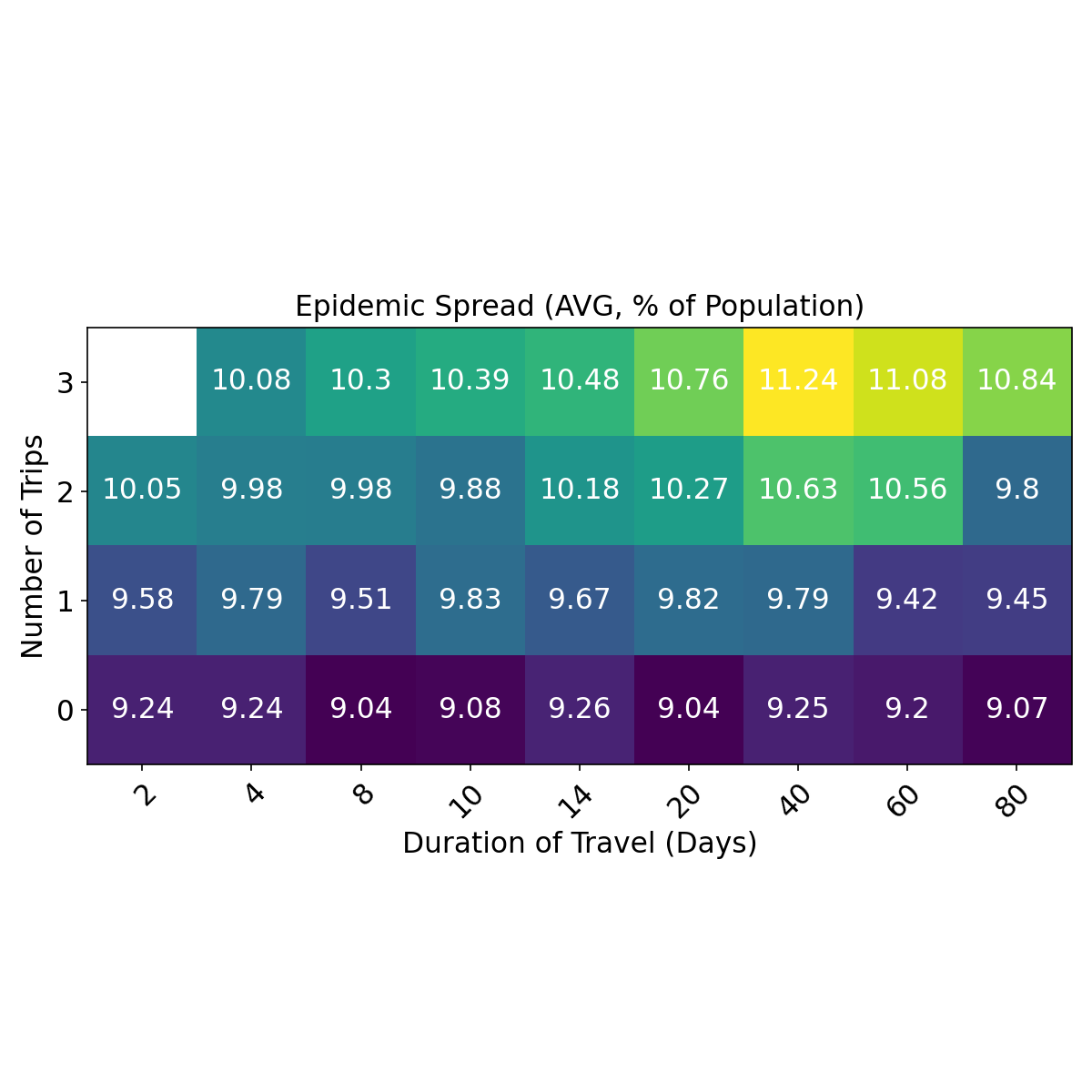}\hfill
    \includegraphics[width=.38\textwidth, trim = 1cm 5cm 0.3cm 6cm, clip]{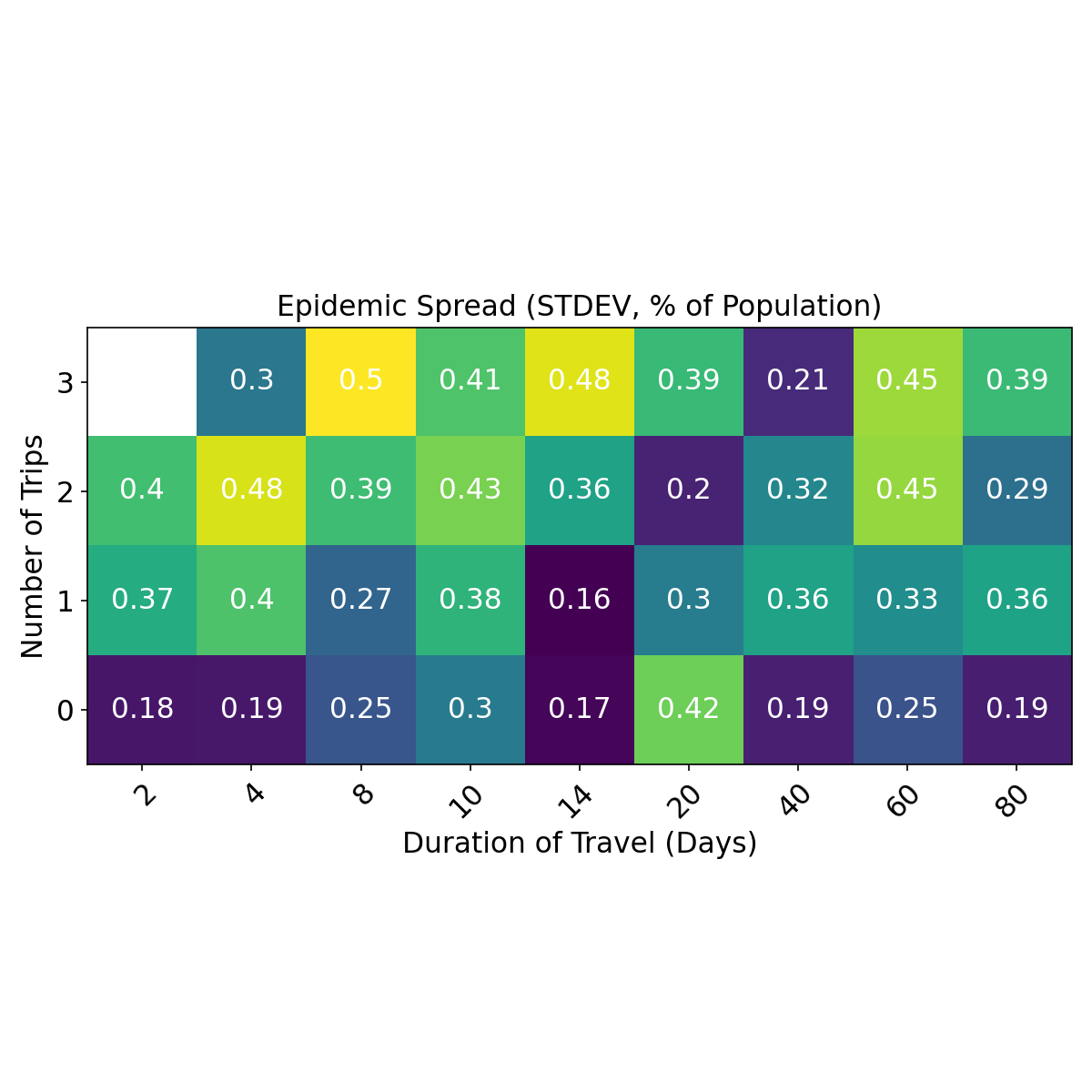}
    \\[\smallskipamount]
    \includegraphics[width=.39\textwidth, trim = 0.2cm 5cm 0.3cm 6cm, clip]{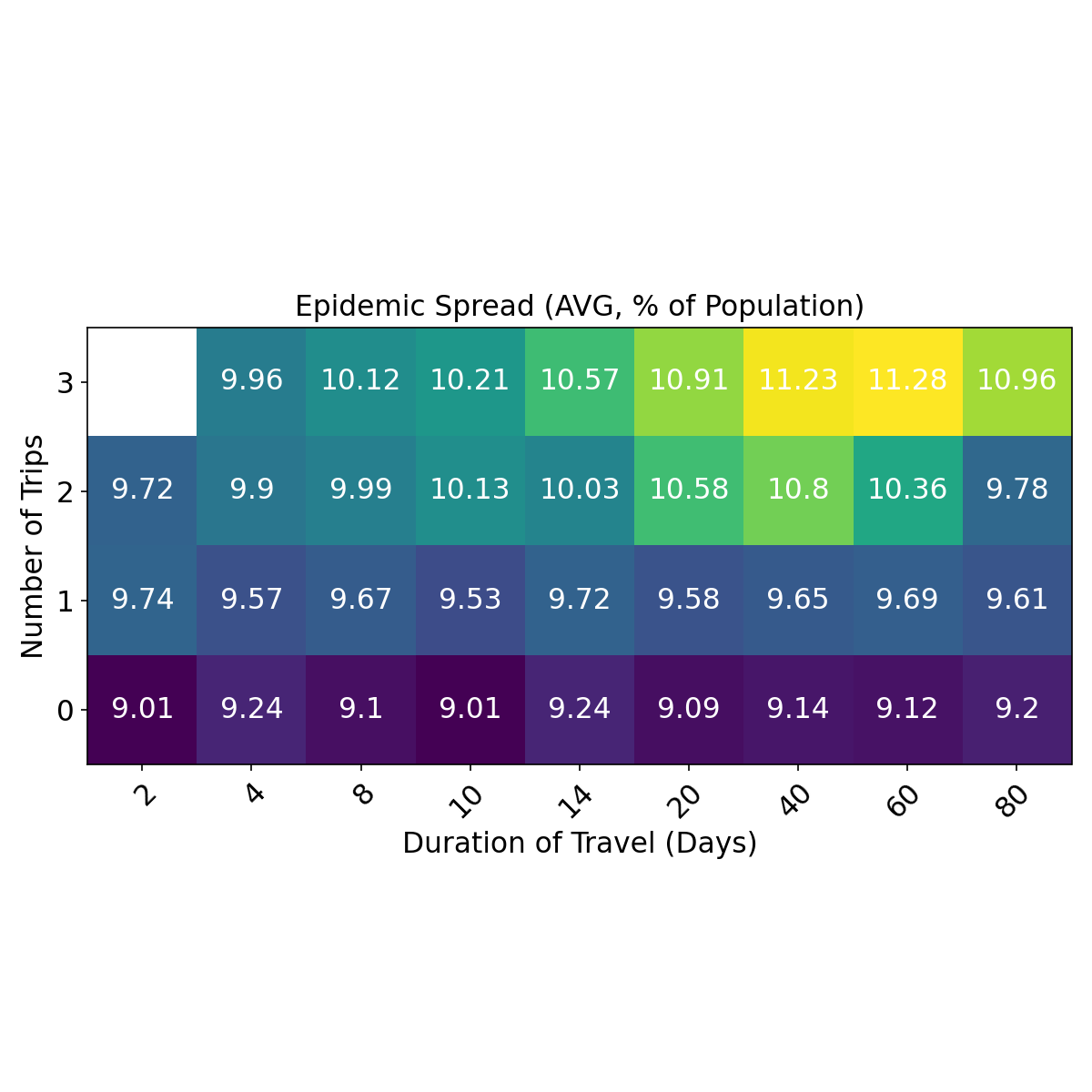}\hfill
    \includegraphics[width=.38\textwidth, trim = 1cm 5cm 0.3cm 6cm, clip]{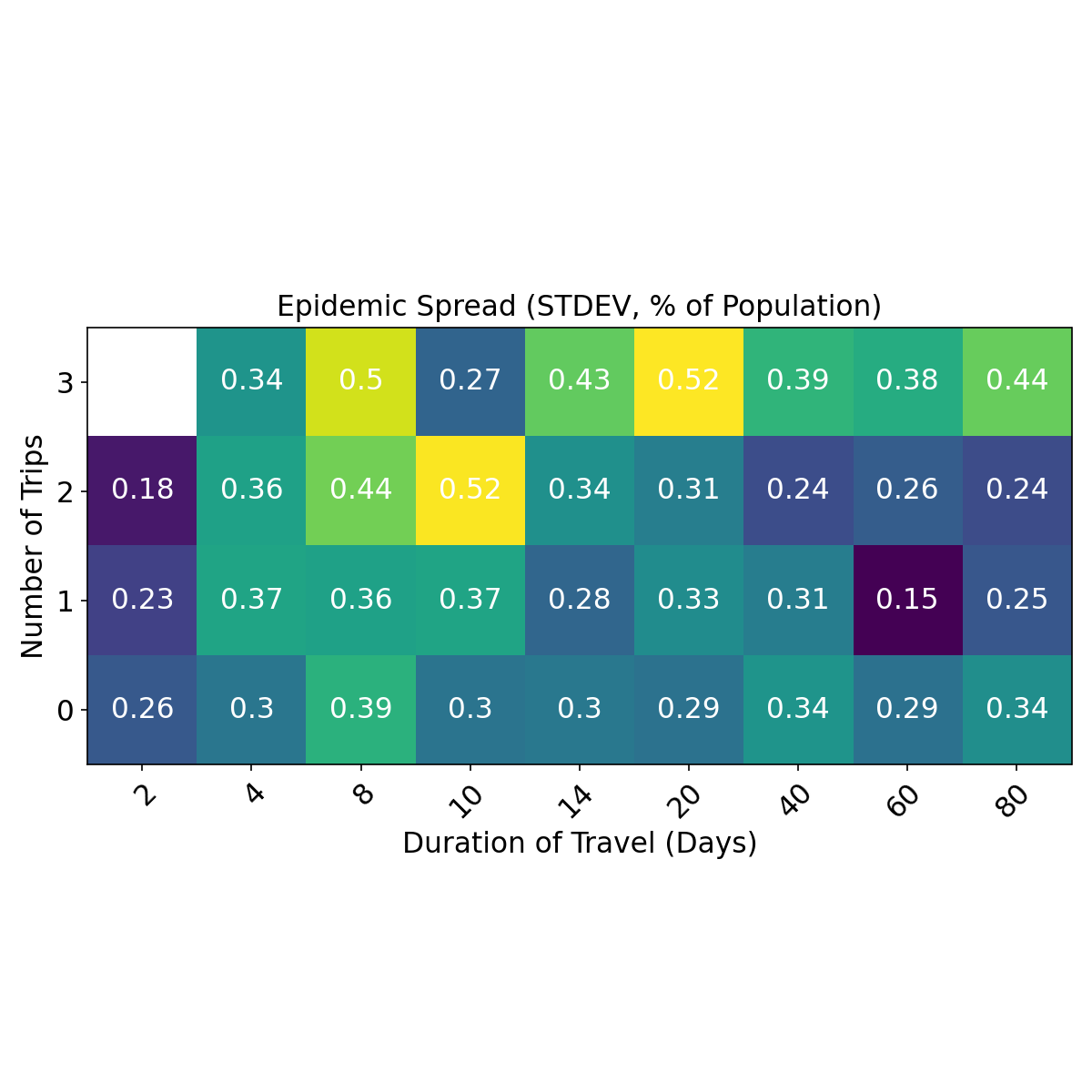}
    \\[\smallskipamount]
    \includegraphics[width=.39\textwidth, trim = 0.2cm 3cm 0.3cm 6cm, clip]{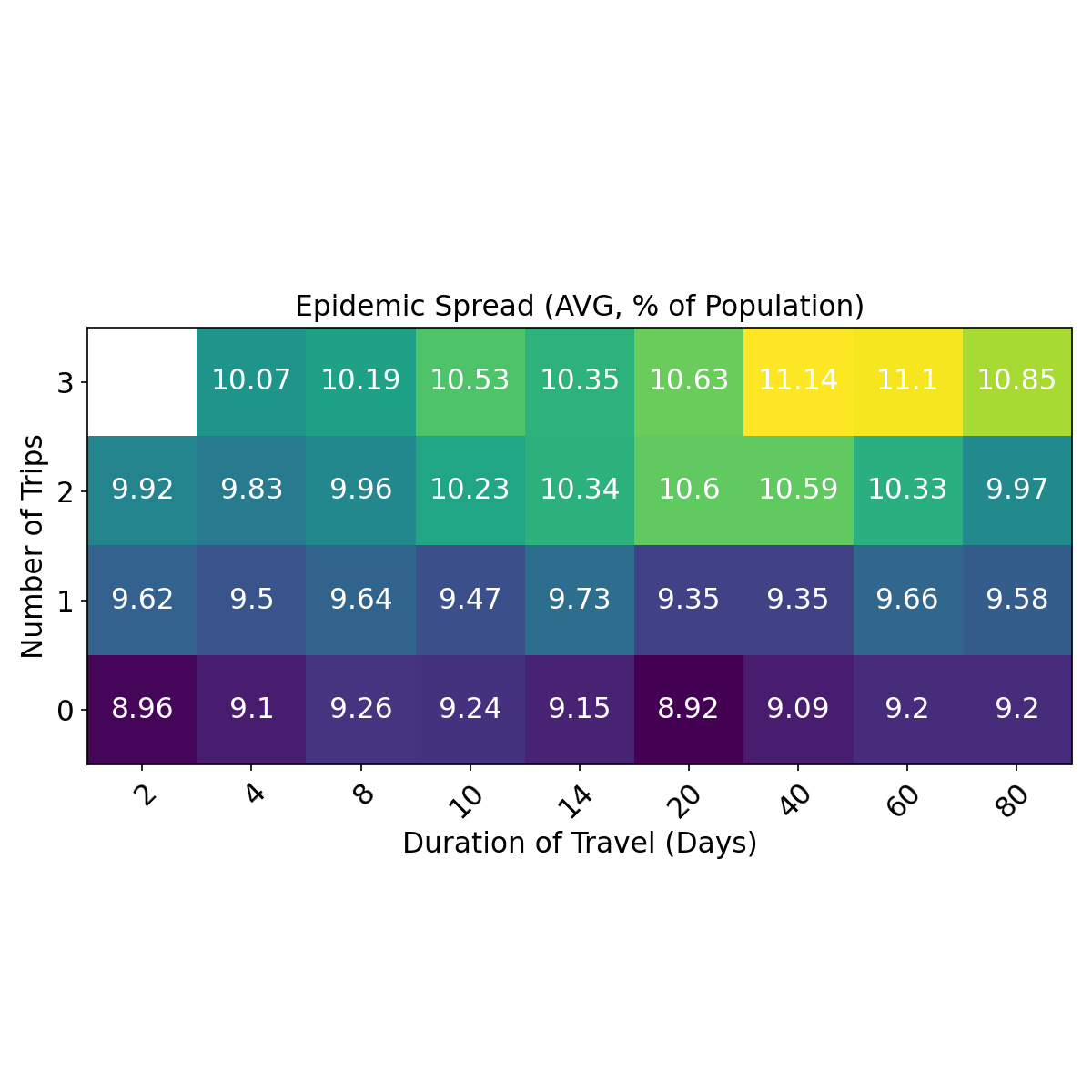}\hfill
    \includegraphics[width=.38\textwidth, trim = 1cm 3cm 0.3cm 6cm, clip]{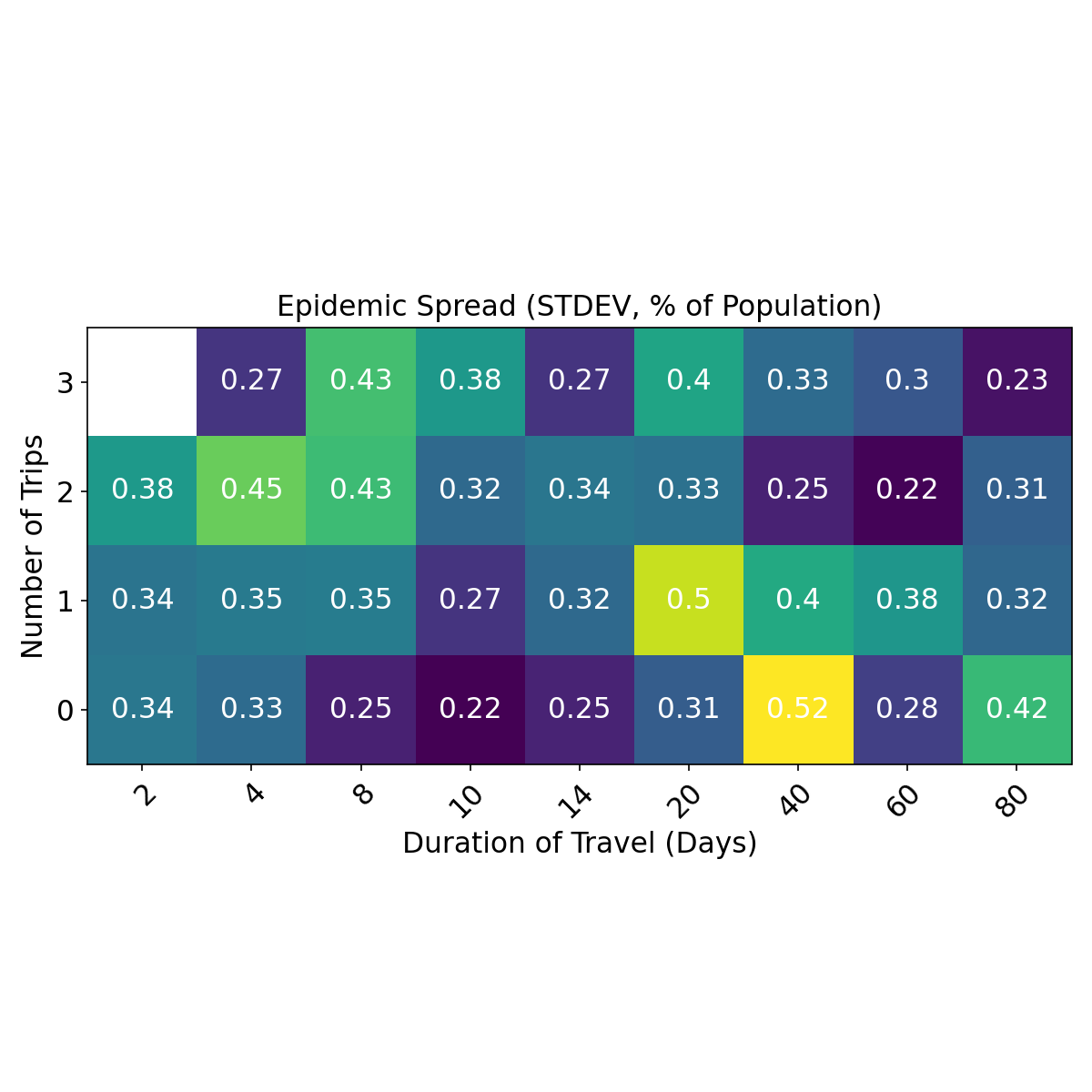}
    \caption{Epidemic spread (number in each cell) as a function of travel cadence (horizontal and vertical axes) when we set different quarantine rates for traveling and non-traveling periods. Six rows correspond to $(q, q_{\mbox{\tiny{travel}}}) = (0, 0), (0, 40\%), (0, 80\%), (40\%, 0), (40\%, 40\%), (40\%, 80\%)$. Averages and standard deviations are summarized from 10 runs.}
    \label{fig:2}
\end{figure}

Figure \ref{fig:1} shows the epidemic spread (number in each cell) as a function of travel cadence (horizontal and vertical axes). In the figure, each row is for a different quarantine level $\in \{0, 10, 20, 30, 40, 50\}\%$. First column shows the average spread over 10 runs. Second column shows the standard deviation of spread over 10 runs. We observe that both the number of trips in a multi-stop travel journey and the total duration of travel impact the overall disease transmission. Worst (most prevalent) transmission happens when people travel multiple stops, with a per-trip duration being equal to or slightly below the mean time to go through exposed and infected states.
Figure \ref{fig:2} shows the epidemic spread (number in each cell) as a function of travel cadence (horizontal and vertical axes) when we set different quarantine rates for traveling ($q_{\mbox{\tiny{travel}}}$) and non-traveling ($q$) periods. In this figure, six rows of subplots correspond to $(q, q_{\mbox{\tiny{travel}}}) = (0, 0), (0, 40\%), (0, 80\%), (40\%, 0), (40\%, 40\%)$, $(40\%, 80\%)$. Averages and standard deviations are summarized from 10 runs. Results from this second set of experiments do not differ qualitatively from the first set - the heat maps in Figure \ref{fig:1} are similar to the corresponding heat maps with equivalent travel cadence tuples in Figure \ref{fig:2}. This is slightly surprising, because it shows that population mixing among travelers \emph{while they are on the move} does not significantly increase disease spread when compared with the importance of quarantine rate for normal, non-travel interactions.

Overall, our simulation results would be most informative for countries or regions that do not have complete control over how travelers interact with ``locals'' during and after their trips. A key takeaway is that, two levers can lead to significant suppression of disease spread: high quarantine rate and reduced travel frequency.
Reducing the travel frequency includes both reducing the number of trips overall, and increasing the wait time between two trips. 
Our simulation results support the rationale, and provide a quantification of the travel restriction strategies adopted by, for example, the \shortciteN{GovernmentofCanada2021}: traveling into Canada is limited to those with immediate ties to the country, and travelers must stay within Canada for 15 days or more before moving onto their next destination.

In contrast to confirming the effectiveness of existing strategies for existing disease (with known transmission characteristics), the arguably more important question is: How should we apply this model to recommend travel guidelines in the face of an unavoidable, novel disease in the future? 
To answer this question, we note that two types of uncertainties complicate the identification of best policy. First, it is difficult and sometimes impossible for everyone to diagnose their infection status when the disease is novel. Therefore it is difficult to precisely apply quarantine on those who are infected. To this end, it is then important for policymakers to encourage reduced interaction among everyone, which essentially is a less precise but still effective way of implementing quarantine (we have simulated other cases with reduced social interaction rate, and the results are consistent).
Second, a novel disease would have unknown exposed and infected duration. Thus, it is not robust to rely on the ``many trips but slow travel'' regime to control spread. The initial focus of governments should be reducing or shutting down travel. When disease transmission characteristics are understood more, local and national governments can then relax the travel guidelines to allow single- or multi-stop travelers that stay in each region for extended periods of time. This also highlights why we need to understand the clinical and public health characteristics of a novel disease rapidly, in order to safely support social, humanitarian, and economic activities during a pandemic.

\section{Conclusion}
In this work, we design and test a simulation model to answer the question: how does travel cadence impact epidemic spread?
We find that travel impacts epidemic spread significantly. 
Our analysis adds to the literature by modeling travel behavior more precisely, and also by examining the long term effect. Previously, works studying travel ban mostly focus on the effect of \emph{delaying} disease transimission in the early stage, and also mostly relying on empirical analysis. Compared with other simulation models such as metapopulation disease transmission models, we provide a higher resolution into the modeling of travel (rather than indirectly influencing the contact rate by defining weights or connectivity parameters as a proxy to represent the contact rate between populations, we directly specify the number of people travelling from one partition to another along with the specific interaction parameters local to that region).
Overall, we found that reducing travel frequency (reducing the number of trips and increasing the duration of stay for each trip during the travel duration) is important. Quarantine rate in general population has a strong impact on the spread. When facing a novel disease with unknown characteristics, it is then important for policymakers to reduce the number of trips, encourage social distancing, and understand the the disease exposed-infected-recovered duration before allowing slow multi-stop travel.


\section*{ACKNOWLEDGMENTS}

This project is funded in part by Carnegie Mellon University's Mobility21 National University Transportation Center, which is sponsored by the US Department of Transportation.

{\footnotesize
\bibliographystyle{wsc}
\bibliography{wsc21}
}

\section*{AUTHOR BIOGRAPHY}

{\footnotesize
\noindent {\bf LAUREN STREITMATTER} is an undergraduate student at the University of Toronto majoring in the Engineering Science Energy Systems Engineering option and minoring in Environmental Engineering. She is currently on a one-year co-op term before returning to the final year of her program. Her research interests include model building and forecasting for applications ranging from disease spread to climate change and sustainable policies. She is interested in attending graduate school to study the intersection of technology, environmental policy, and social justice or to pursue an advanced degree in renewable energy and environmental engineering. Her email address is \href{mailto:lauren.streitmatter@mail.utoronto.ca}{lauren.streitmatter@mail.utoronto.ca}.\\

\noindent {\bf PETER ZHANG} is an assistant professor in operations research at the Heinz College of Information Systems and Public Policy, Carnegie Mellon University. He received Bachelor of Applied Science in Engineering Science and Master of Applied Science from the University of Toronto, and PhD in Engineering Systems from the Institute of Data, Systems, and Society at Massachusetts Institute of Technology. His research interests are primarily in the theory of robust optimization, dynamic decision making under uncertainty, and applications in socio-technical systems such as transportation, supply chain, and machine learning. His email address is \href{mailto:pyzhang@cmu.edu}{pyzhang@cmu.edu}, and his website is \href{http://www.andrew.cmu.edu/user/yunz2/}{http://www.andrew.cmu.edu/user/yunz2/}.
}

\end{document}

%% file: wscbib.tex
\makeatletter
\let\@internalcite\cite
\def\cite{\def\@citeseppen{-1000}%
    \def\@cite##1##2{(##1\if@tempswa , ##2\fi)}%
    \def\citeauthoryear##1##2##3{##1 ##3}\@internalcite}
\def\citeNP{\def\@citeseppen{-1000}%
    \def\@cite##1##2{##1\if@tempswa , ##2\fi}%
    \def\citeauthoryear##1##2##3{##1 ##3}\@internalcite}
\def\citeN{\def\@citeseppen{-1000}%
    \def\@cite##1##2{##1\if@tempswa, ##2)\else{}\fi}%
    \def\citeauthoryear##1##2##3{##1 (##3)}\@citedata}
\def\citeA{\def\@citeseppen{-1000}%
    \def\@cite##1##2{(##1\if@tempswa , ##2\fi)}%
    \def\citeauthoryear##1##2##3{##1}\@internalcite}
\def\citeANP{\def\@citeseppen{-1000}%
    \def\@cite##1##2{##1\if@tempswa , ##2\fi}%
    \def\citeauthoryear##1##2##3{##1}\@internalcite}
\def\shortcite{\def\@citeseppen{-1000}%
    \def\@cite##1##2{(##1\if@tempswa , ##2\fi)}%
    \def\citeauthoryear##1##2##3{##2 ##3}\@internalcite}
\def\shortciteNP{\def\@citeseppen{-1000}%
    \def\@cite##1##2{##1\if@tempswa , ##2\fi}%
    \def\citeauthoryear##1##2##3{##2 ##3}\@internalcite}
\def\shortciteN{\def\@citeseppen{-1000}%
    \def\@cite##1##2{##1\if@tempswa, ##2\else{}\fi}%
    \def\citeauthoryear##1##2##3{##2 (##3)}\@citedata}
\def\shortciteA{\def\@citeseppen{-1000}%
    \def\@cite##1##2{(##1\if@tempswa , ##2\fi)}%
    \def\citeauthoryear##1##2##3{##2}\@internalcite}
\def\shortciteANP{\def\@citeseppen{-1000}%
    \def\@cite##1##2{##1\if@tempswa , ##2\fi}%
    \def\citeauthoryear##1##2##3{##2}\@internalcite}
\def\citeyear{\def\@citeseppen{-1000}%
    \def\@cite##1##2{(##1\if@tempswa , ##2\fi)}%
    \def\citeauthoryear##1##2##3{##3}\@citedata}
\def\citeyearNP{\def\@citeseppen{-1000}%
    \def\@cite##1##2{##1\if@tempswa , ##2\fi}%
    \def\citeauthoryear##1##2##3{##3}\@citedata}
%
%
%
\def\@citedata{%
    \@ifnextchar [{\@tempswatrue\@citedatax}%
                  {\@tempswafalse\@citedatax[]}%
}

\def\@citedatax[#1]#2{%
\if@filesw\immediate\write\@auxout{\string\citation{#2}}\fi%
  \def\@citea{}\@cite{\@for\@citeb:=#2\do%
    {\@citea\def\@citea{, }\@ifundefined
       {b@\@citeb}{{\bf ?}%
       \@warning{Citation `\@citeb' on page \thepage \space undefined}}%
{\csname b@\@citeb\endcsname}}}{#1}}%

%
\def\@citex[#1]#2{%
\if@filesw\immediate\write\@auxout{\string\citation{#2}}\fi%
  \def\@citea{}\@cite{\@for\@citeb:=#2\do%
    {\@citea\def\@citea{; }\@ifundefined
       {b@\@citeb}{{\bf ?}%
       \@warning{Citation `\@citeb' on page \thepage \space undefined}}%
{\csname b@\@citeb\endcsname}}}{#1}}%

%
\def\@biblabel#1{}
\makeatother



\newdimen\bibindent
\bibindent=0.0em
\def\thebibliography#1{\section*{\refname}\list
   {}{\settowidth\labelwidth{[#1]}
   \leftmargin\parindent
   \itemindent -\parindent
   \listparindent \itemindent
   \itemsep 0pt
   \parsep 0pt}
   \def\newblock{}
   \sloppy
   \sfcode`\.=1000\relax}

%% file: wsc21.bbl
\begin{thebibliography}{}

\bibitem[\protect\citeauthoryear{Adekunle, Meehan, Rojas‐Alvarez, Trauer, and
  McBryde}{Adekunle et~al.}{2020}]{Adekunle2020}
Adekunle, A., M.~Meehan, D.~Rojas‐Alvarez, J.~Trauer, and E.~McBryde. 2020,
  aug.
\newblock ``{Delaying the COVID‐19 epidemic in Australia: evaluating the
  effectiveness of international travel bans}''.
\newblock {\em Australian and New Zealand Journal of Public
  Health\/}~44(4):257--259.


\bibitem[\protect\citeauthoryear{Anzai, Kobayashi, Linton, Kinoshita, Hayashi,
  Suzuki, Yang, Jung, Miyama, Akhmetzhanov, and Nishiura}{Anzai
  et~al.}{2020}]{Anzai2020}
Anzai, A., T.~Kobayashi, N.~M. Linton, R.~Kinoshita, K.~Hayashi, A.~Suzuki,
  Y.~Yang, S.-m. Jung, T.~Miyama, A.~R. Akhmetzhanov, and H.~Nishiura. 2020,
  feb.
\newblock ``{Assessing the Impact of Reduced Travel on Exportation Dynamics of
  Novel Coronavirus Infection (COVID-19)}''.
\newblock {\em Journal of Clinical Medicine\/}~9(2):601.


\bibitem[\protect\citeauthoryear{Apolloni, Poletto, Ramasco, Jensen, and
  Colizza}{Apolloni et~al.}{2014}]{Apolloni2014}
Apolloni, A., C.~Poletto, J.~J. Ramasco, P.~Jensen, and V.~Colizza. 2014.
\newblock ``{Metapopulation epidemic models with heterogeneous mixing and
  travel behaviour}''.
\newblock {\em Theoretical Biology and Medical Modelling\/}~11(1):3.


\bibitem[\protect\citeauthoryear{Arino and van~den Driessche}{Arino and van~den
  Driessche}{2003}]{Arino2003}
Arino, J., and P.~van~den Driessche. 2003, jan.
\newblock ``{A multi-city epidemic model}''.
\newblock {\em Mathematical Population Studies\/}~10(3):175--193.


\bibitem[\protect\citeauthoryear{Ball, Britton, House, Isham, Mollison, Pellis,
  and {Scalia Tomba}}{Ball et~al.}{2015}]{Ball2015}
Ball, F., T.~Britton, T.~House, V.~Isham, D.~Mollison, L.~Pellis, and
  G.~{Scalia Tomba}. 2015, mar.
\newblock ``{Seven challenges for metapopulation models of epidemics, including
  households models}''.
\newblock {\em Epidemics\/}~10:63--67.


\bibitem[\protect\citeauthoryear{Birge, Candogan, and Feng}{Birge
  et~al.}{2020}]{Birge2020}
Birge, J.~R., O.~Candogan, and Y.~Feng. 2020.
\newblock ``{Controlling Epidemic Spread: Reducing Economic Losses with
  Targeted Closures}''.
\newblock {\em SSRN Electronic Journal\/}.


\bibitem[\protect\citeauthoryear{Chinazzi, Davis, Ajelli, Gioannini, Litvinova,
  Merler, {Pastore y Piontti}, Mu, Rossi, Sun, Viboud, Xiong, Yu, Halloran,
  Longini, and Vespignani}{Chinazzi et~al.}{2020}]{Chinazzi2020}
Chinazzi, M., J.~T. Davis, M.~Ajelli, C.~Gioannini, M.~Litvinova, S.~Merler,
  A.~{Pastore y Piontti}, K.~Mu, L.~Rossi, K.~Sun, C.~Viboud, X.~Xiong, H.~Yu,
  M.~E. Halloran, I.~M. Longini, and A.~Vespignani. 2020, apr.
\newblock ``{The effect of travel restrictions on the spread of the 2019 novel
  coronavirus (COVID-19) outbreak}''.
\newblock {\em Science\/}~368(6489):395--400.


\bibitem[\protect\citeauthoryear{Costantino, Heslop, and MacIntyre}{Costantino
  et~al.}{2020}]{Costantino2020}
Costantino, V., D.~J. Heslop, and C.~R. MacIntyre. 2020, aug.
\newblock ``{The effectiveness of full and partial travel bans against COVID-19
  spread in Australia for travellers from China during and after the epidemic
  peak in China}''.
\newblock {\em Journal of Travel Medicine\/}~27(5).


\bibitem[\protect\citeauthoryear{{Government of Canada}}{{Government of
  Canada}}{2021}]{GovernmentofCanada2021}
{Government of Canada} 2021.
\newblock ``{Coronavirus disease (COVID-19): Who can travel to Canada –
  Citizens, persons registered under Canada's Indian Act, permanent residents,
  foreign nationals and refugees}''.

\bibitem[\protect\citeauthoryear{Mangrum and Niekamp}{Mangrum and
  Niekamp}{2020}]{Mangrum2020}
Mangrum, D., and P.~Niekamp. 2020, dec.
\newblock ``{JUE Insight: College student travel contributed to local COVID-19
  spread}''.
\newblock {\em Journal of Urban Economics\/}:103311.


\bibitem[\protect\citeauthoryear{Ni and Weng}{Ni and Weng}{2009}]{Ni2009}
Ni, S., and W.~Weng. 2009, jan.
\newblock ``{Impact of travel patterns on epidemic dynamics in heterogeneous
  spatial metapopulation networks}''.
\newblock {\em Physical Review E\/}~79(1):016111.


\bibitem[\protect\citeauthoryear{Watts, Muhamad, Medina, and Dodds}{Watts
  et~al.}{2005}]{Watts2005}
Watts, D.~J., R.~Muhamad, D.~C. Medina, and P.~S. Dodds. 2005, aug.
\newblock ``{Multiscale, resurgent epidemics in a hierarchical metapopulation
  model}''.
\newblock {\em Proceedings of the National Academy of
  Sciences\/}~102(32):11157--11162.


\bibitem[\protect\citeauthoryear{You, Deng, Hu, Sun, Lin, Zhou, Pang, Zhang,
  Chen, and Zhou}{You et~al.}{2020}]{You2020}
You, C., Y.~Deng, W.~Hu, J.~Sun, Q.~Lin, F.~Zhou, C.~H. Pang, Y.~Zhang,
  Z.~Chen, and X.-H. Zhou. 2020, jul.
\newblock ``{Estimation of the time-varying reproduction number of COVID-19
  outbreak in China}''.
\newblock {\em International Journal of Hygiene and Environmental
  Health\/}~228:113555.


\end{thebibliography}
